\documentclass[graybox]{svmult}

\usepackage{mathptmx}       
\usepackage{helvet}         
\usepackage{courier}        
\usepackage{type1cm}        
\usepackage{makeidx}         
\usepackage{graphicx}        
\usepackage{multicol}        
\usepackage[bottom]{footmisc}
\usepackage{textcomp}
\usepackage{gensymb}
\usepackage{natbib}

\newcommand{\kms}{\mbox{$\>{\rm km\, s^{-1}}$}}

\newcommand{\kpc}{\mbox{$\>{\rm kpc}$}} 
\newcommand{\Gyr}{\mbox{$\>{\rm Gyr}$}}
\newcommand{\Myr}{\mbox{$\>{\rm Myr}$}}
\newcommand{\Msun}{\mbox{${M_{\odot}}$}}
\newcommand{\Rd}{\mbox{$R_{\rm d}$}}

\newcommand\degrees{^\circ}
\newcommand\hi{{\sc Hi}}

\newcommand{\sig}[1]{\mbox{$\sigma_{#1}$}}
\newcommand{\om}[1]{\mbox{$\Omega_{#1}$}}
\newcommand{\feh}{\mbox{$\rm [Fe/H]$}}

\newcommand{\alfe}{\mbox{$\rm [\alpha/Fe]$}}

\def\spose#1{\hbox to 0pt{#1\hss}}
\def\gtsim{\mathrel{\spose{\lower.5ex \hbox{$\mathchar"218$}}
     \raise.4ex\hbox{$\mathchar"13E$}}}
\def\ltsim{\mathrel{\spose{\lower.5ex\hbox{$\mathchar"218$}}
     \raise.4ex\hbox{$\mathchar"13C$}}}

\def\ie{{i.e.,}}

\makeindex             

\begin{document}

\title*{The Impact of Stellar Migration on disk Outskirts}
\author{Victor P. Debattista, Rok Ro\v skar, Sarah R. Loebman}
\institute{Victor P. Debattista \at Jeremiah Horrocks Institute, University of Central Lancashire, Preston, PR1 2HE, UK, \email{vpdebattista@uclan.ac.uk} 
  \and Rok Ro\v skar \at Research Informatics, Scientific IT Services, ETH Z\"urich, Weinbergstrasse 11, CH-8092, Z\"urich, Switzerland, \email{rokroskar@gmail.com} 
  \and Sarah R. Loebman \at Department of Astronomy, University of Michigan, 1085 S. University Ave., Ann Arbor, MI, 48109, USA, \email{sloebman@umich.edu}
}
%
%
\maketitle

\abstract{Stellar migration, whether due to trapping by transient
  spirals (churning), or to scattering by non-axisymmetric
  perturbations, has been proposed to explain the presence of stars in
  outer disks.  After a review of the basic theory, we present
  compelling, but not yet conclusive, evidence that churning has been
  important in the outer disks of galaxies with type~II (down-bending)
  profiles, while scattering has produced the outer disks of type~III
  (up-bending) galaxies.  In contrast, field galaxies with type~I
  (pure exponential) profiles appear to not have experienced
  substantial migration.  We conclude by suggesting work that would
  improve our understanding of the origin of outer disks.  }

\section{Introduction}
\label{sec:intro}

By the mid-90s the view of disks that prevailed was that they are
truncated at some radius with little or no stars at larger radii.
Using photographic images of three edge-on galaxies, \citet{vdkruit79} had
shown that disks appear sharply truncated.  Based on four edge-on
galaxies, \citet{vanderkruit_searle82} measured truncation
\index{truncation} radii at $4.2 \pm 0.6\,\Rd$, where \Rd\ is the
scalelength of the disk.  In face-on galaxies, \citet{vdkruit88} found
closely spaced light contours in the outer disks, which he interpreted
as the advent of the truncation.  \citet{vdkruit87} suggested these
truncations are the result of detailed angular momentum conservation
in the collapse of a uniformly rotating gas sphere.  While \hi\ disks
are typically larger than the break radius, \citet{vdkruit07} showed
that \hi\ warps \index{warps} often begin at about the truncation
radius.  \citet{kregel+02} found truncations in 20 of 34 galaxies with
a slightly smaller truncation radius, $3.6\pm 0.6\,\Rd$.
\citet{pohlen+00}, on the basis of a sample of 31 edge-on galaxies,
revised the mean truncation radius further to $2.9 \pm 0.7 \,\Rd$
(truncations at $11-35\,\kpc$).

Then the picture of disk profiles started changing.  Already in
NGC~4565 (\citealt{naeslund_joersaeter97}) and IC~5249 (\citealt{byun98}) a
transition from one exponential profile to a steeper one, rather than
a sharp truncation, had been found.  In his sample of four edge-on
galaxies, \citet{degrijs+01} found that truncations are not perfectly
sharp.  Finally the deep optical imaging of 72 edge-on
(\citealt{pohlen_phd}) and three face-on (\citealt{pohlen+02}) galaxies firmly
established that disk galaxies have broken-exponential
profiles\index{broken-exponential profiles}, rather than sharp
truncations.  The transition between the two exponentials has come to
be known as a ``break'', although the term ``truncation'' persists.  This
discovery led to increased interest in disk outskirts because the
presence of stars at large radii needed an explanation.  The first
workshop dedicated to disk outskirts, ``Outer edges of disk galaxies:
A truncated perspective'', was held at the Lorentz Center in 2005.
That same year \citet{erwin+05} identified a new class of disk
profiles, the type~III ``anti-truncated'' profiles\index{anti-truncated
  profiles}.  The first pure $N$-body simulations aimed at producing
disk profiles with breaks were presented by \citet{debattista+06}, who
proposed angular momentum redistribution during bar formation as the
cause, while \citet{bournaud+07} and \citet{foyle+08} included also the
effect of gas.  \citet{roskar+08a} introduced the idea that disk
outskirts are populated by stars that have migrated outwards without
heating.

A decade later much progress has been made but much more remains to be
understood.  While generally comprising only a small fraction of the
overall stellar mass, disk outskirts are an ideal laboratory for
understanding the processes of galaxy formation.  These are regions
where the dark matter dominates and that are therefore easily perturbed,
are generally inhospitable to star formation, and are the interface at
which gas, and some satellites, are accreted.
In recent years it has become clear that the stars in these regions
are generally old.  Are these stars the product of ongoing inefficient
star formation, or the fossil of a long-ago, efficient, in-situ star
formation?  Was mass migrated to the outer disk (by bars, clumps,
or spirals)?  Were the stars accreted from satellites?  Understanding
which of these processes dominates, and where, provides crucial
information on how galaxies formed.
This review explores the extent to which stellar migration is
responsible for populating the outer disks.  In the remainder of this
Section we clarify what we mean by profile breaks.  In Sect.~\ref{sec:demographics} we review the demographics of different profile
types.  We introduce stellar migration and angular momentum
redistribution in Sect.~\ref{sec:migration}.  We then consider each
of the profile types in turn in Sect.~\ref{sec:typeII} (type~IIs),
\ref{sec:typeI} (type~Is), and \ref{sec:typeIII} (type~IIIs).  Section~\ref{sec:future} concludes with suggestions for future directions that
can greatly advance our understanding of the origin of stars in outer
disks.

\subsection{Our Definition of Breaks}

In the truncated picture of disks, the meaning and location of a
truncation is unambiguous. Instead disks do not end abruptly but
extend significantly beyond a break.  While disks are often well fit
by piecewise exponential profiles, the presence of bars, lenses,
spirals, rings and asymmetries add bumps and wiggles to profiles.  For
instance, \citet{laine+14} noted that their galaxies included breaks
associated with lenses, with rings, and with star formation breaks.
Therefore sometimes it is difficult to assign the location of breaks,
particularly if the data being fit are shallow.  Such difficulties
account for the significant differences in the locations of breaks
assigned by different authors.  For example, in NGC~4244 the resolved
stellar population study of \citet{dejong+07} found a very clear break
at $420''$ while \citet{martin-navarro+12} found one at $\sim 290''$
and another at $\sim 550''$.  Although sharp truncations are not
supported by observations, star formation is often observed to be more
strongly truncated than the overall mass distribution
(\citealt{bakos+08}).  How stars end up past this point is a very
important question that gives us insight into important processes that
operate in most disk galaxies.  With this in mind, this review focuses
not on the bumps and wiggles produced by bars, rings, or spirals, but
on breaks related to star formation.  Thus our primary focus is on
type~II profiles, which exhibit such breaks, whereas our interest in
pure exponential (type~I) profiles comes from the absence of these
breaks, and in anti-truncated (type~III) profiles from their ability to
populate the outer disk via some other mechanism.


\section{Demographics of Profile Type\index{Demographics of profile type}}
\label{sec:demographics}

Stellar disk profiles are piece-wise exponential\footnote{Bulges at the
  centres of galaxies contribute additional light above an
  exponential.}.  This does not imply that the profile of star
formation itself must be exponential since even a radially constant
star formation rate produces an exponential disk
(e.g., \citealt{roskar+08a}).  Why nature prefers exponential disks is
uncertain, although many studies have attacked this problem from a
variety of perspectives.  Whatever the reason, the exponential profile
is evidently an attractor.
\citet{freeman70} identified two types of disk profiles.  The first
(type~I profile)\index{type~I profile} is exponential to the last
measured point, while in the second (type~II profile)\index{type~II profile} an inner exponential is followed by a steeper one in the
outer disk.  \citet{erwin+05} extended this classification to include
type~III profiles\index{type~III profile}, where the outer profile is
shallower than the inner one.

\citet{pohlen_trujillo06} found that $\sim 60\%$ of late-type spirals have a
type~II profile with the break at $2.5 \pm 0.6 \,\Rd$ (at $\mu_r \sim
23.5 \pm 0.8$ mag\, arcsec$^{-2}$), while $30\%$ have a type~III
profile, with the break at larger radii, $4.9 \pm 0.6 \,\Rd$ (at $\mu_r
\sim 24.7 \pm 0.8$ mag\,arcsec$^{-2}$).  The mix of profile types
varies with Hubble type, with type~IIs more frequent in late-type
galaxies while type~IIIs are more common in early-type galaxies.
\citet{erwin+08} found a breakdown of $27\%:42\%:27\%$ for
types~I~:~II~:~III for 66 early-type galaxies, with $\sim 6\%$ having a
composite type~II+III profile\index{composite type~II+III profile}: a type~II profile at small radii and a type~III further out.
\citet{gutierrez+11}, studying 47 face-on early-type unbarred galaxies,
found that the type~II profiles are more common in late-type galaxies
while the type~I and III profiles are more common in early-types.  By
combining all the published studies and accounting for differences in
sample definitions, they found global (S0-Sd) frequencies of $21\% :
50\% : 38\%$ for types~I~:~II~:~III, with 8\% of galaxies having composite
II+III profiles.  Table \ref{tab:demographics} presents the fraction
of each profile type (with type~II+III profiles counted twice) from
the combination of these three studies.

\begin{table}
  \begin{center}
    \begin{tabular}{| l | c | c | c |}
      \hline
      \multicolumn{1}{|l|}{} &
      \multicolumn{1}{c|}{I} &
      \multicolumn{1}{c|}{II} &
      \multicolumn{1}{c|}{III} \\ 
      \hline
S0-Sb galaxies & & &  \\
 SA &  $0.26 \pm 0.07$ & $0.14 \pm 0.06$ & $0.69 \pm 0.08$ \\
SAB &  $0.18 \pm 0.07$ & $0.45 \pm 0.09$ & $0.48 \pm 0.09$ \\
 SB &  $0.36 \pm 0.07$ & $0.49 \pm 0.07$ & $0.20 \pm 0.06$ \\
 \hline

Sbc-Sd galaxies & & &  \\
 SA &  $0.17 \pm 0.08$ & $0.52 \pm 0.10$ & $0.30 \pm 0.10$ \\
SAB &  $0.12 \pm 0.07$ & $0.67 \pm 0.10$ & $0.21 \pm 0.08$ \\
 SB &  $0.09 \pm 0.06$ & $0.83 \pm 0.08$ & $0.09 \pm 0.06$ \\

 \hline
    \end{tabular}
  \end{center}
  \caption[]{The fraction of type~I, II and III profiles amongst
    unbarred (SA), weakly barred (SAB) and barred (SB) galaxies.  Data
    for S0-Sb galaxies are from \citet{erwin+08} and
    \citet{gutierrez+11}.  The data for Sbc-Sd galaxies are from
    \citet{pohlen_trujillo06}
\label{tab:demographics} }
\end{table}

\subsection{The Role of Environment\index{environment}}
\label{ssec:environment}

\citet{erwin+12} and \citet{roediger+12} found a strong dichotomy
between field and Virgo Cluster lenticulars\index{cluster lenticulars}.  Both studies found that the fraction of type~I
profiles rises at the expense of type~II profiles.
\citet{gutierrez+11} found type~I profiles in one third of lenticular
galaxies, with type~II profiles uncommon amongst them.
\citet{pranger+16} used the Sloan Digital Sky Survey to construct field and cluster samples of
$\sim 100$ galaxies each, in a narrow mass range ($1-4\times
10^{10}\,\Msun$).  They found that type~I galaxies are three times more
frequent in clusters than in the field.
\citet{maltby+12a} came to a very different conclusion, finding roughly
half of both their field and cluster samples having a type~I profile
and only $10\%$ having a type~II profile.  However, because they
concentrated on outer disks through a surface brightness cut at $24 <
\mu < 26.5$ mag\,arcsec$^{-2}$, they missed most type~II breaks, which
are typically brighter.  Nevertheless, their fraction of type~III
profiles is consistent with that of the field sample of
\citet{pohlen_trujillo06}.

\subsection{Redshift Evolution\index{redshift evolution}}

\begin{figure}[]
  \sidecaption
\centerline{  
\includegraphics[angle=0.,width=\hsize]{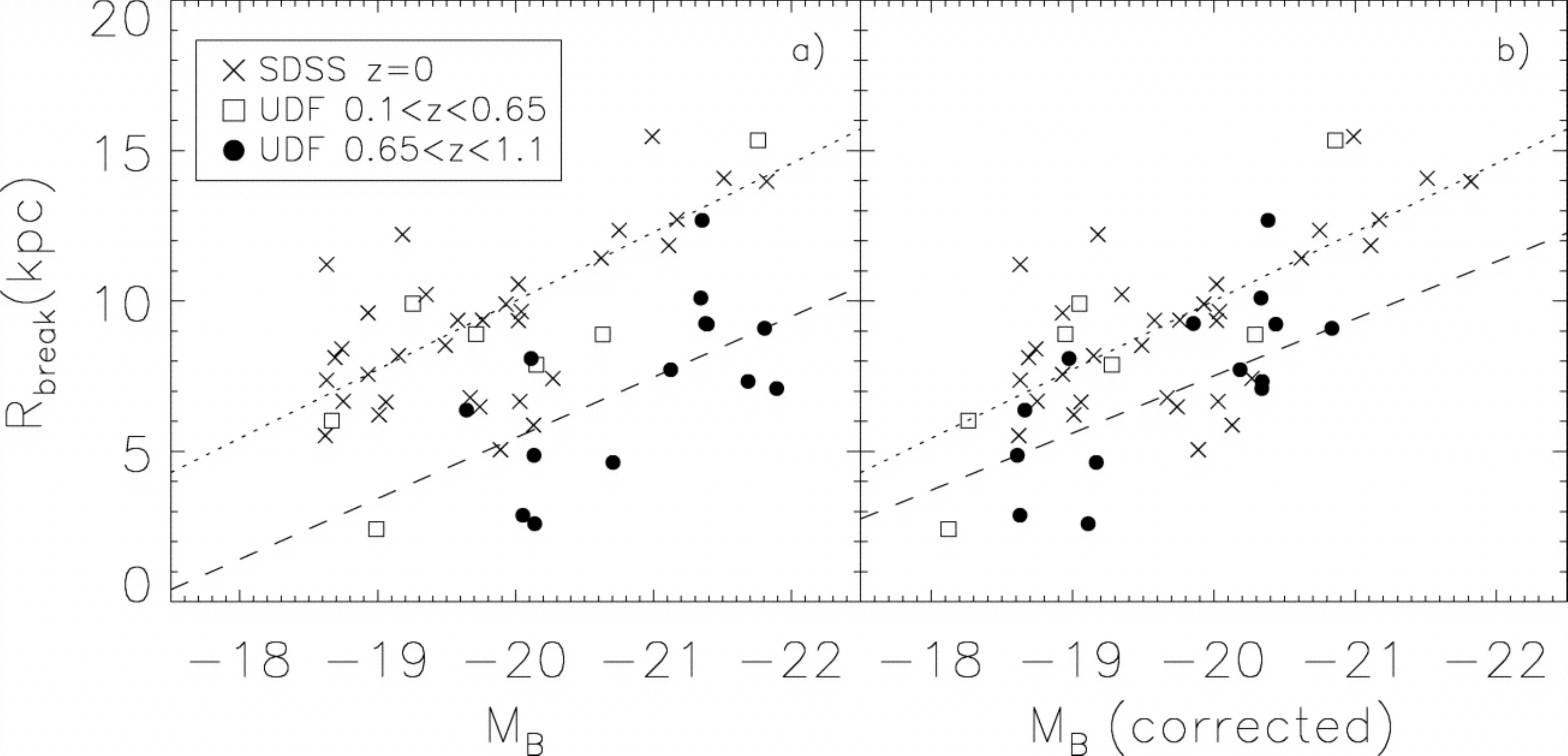}}
\caption{Break radius as a function of rest-frame $B$-band absolute
  magnitude for galaxies at different redshift, as indicated in the
  inset at left.  ({\it a}) The observed relation.  ({\it b}) The relation after
  correcting the absolute magnitude for the mean surface brightness
  evolution since $z=1$.  The dashed lines show the best-fitting
  relation for the distant sample while the dotted line shows the
  local relation.  Reproduced with permission from
  \citet{trujillo_pohlen05}  }
\label{fig:tp05} 
\end{figure}

\citet{perez04} found six breaks in a sample of 16 galaxies at $0.6
\leq z \leq 1.0$; the average break radius was $1.8 \,\Rd$, smaller than
in local samples although the detection was biased to smaller radii.
In a study of 36 galaxies to $z \sim 1$, shown in Fig.~\ref{fig:tp05},
\citet{trujillo_pohlen05} found that the location of the breaks,
corrected for the mean surface brightness evolution, has increased by
1-3\,\kpc\ ($25\%$), while the $\sim 10\times$ larger sample of
\citet{azzollini+08b} found that, at fixed mass, the break radius has
increased by a factor $1.3 \pm 0.1$ since $z=1$.


\section{Stellar Migration\index{migration}}
\label{sec:migration}

The breaks in disk profiles are generally at $2-5\,\Rd$
(\citealt{pohlen_trujillo06}).  At these large radii, the surface density
is low and the total mass in the outer disk is small.  \citet{bakos+08}
found that the mass of stars in the outer disk constitutes $14.7\%\pm
1.2\%$ in type~II galaxies and $9.2\% \pm 1.4\%$ in type~III galaxies.
Therefore even a quite small transfer of mass from small to large
radii can have a significant effect on the overall profile shape.  For
a pure exponential profile, the total mass outside $4\,\Rd$ is less than
$10\%$ of the total.  If just one quarter of the mass between $\Rd$
and $2\,\Rd$ (\ie\ $\sim 8\%$ of the total disk mass) is redistributed
to radii outside $4\,\Rd$, this nearly doubles the mass at those radii.
This means that a galaxy does not need substantial mass redistribution
in order for its outer profile to be significantly
altered.\footnote{This also poses a challenge for simulations studying
  the evolution of disk outskirts, which need to minimise noise; for
  instance initial conditions that are even slightly out of
  equilibrium can significantly alter the profile at large radii.  The
  most widely used technique for producing initial conditions for
  simulations assumes Gaussian velocity distributions, which is only
  an approximate equilibrium.  \citet{kazantzidis+04} presented a
  spectacular example of how this approximation leads to dramatically
  wrong conclusions in the case of satellite galaxy disruption.  The
  freely available {\sc GalactICS} initial conditions code
  (\citealt{kuijken_dubinski95, widrow_dubinski05, widrow+08}) instead
  uses the correct distribution function method for setting up
  galaxies.  Likewise, the {\sc Galaxy} package (\citealt{sellwood14})
  includes software for setting up disk galaxies in proper
  equilibrium.}

\subsection{Migration via Transient Spirals}

The energy in a rotating frame, the Jacobi energy\index{Jacobi
  energy}, $E_{\rm J}$, is a conserved quantity:
\begin{equation}
E_{\rm J} = E - \om{\rm p} L_z,
\end{equation}
where $E$ and $L_z$ are the energy and angular momenta in the inertial
frame, and \om{\rm p}\ is the angular frequency of the rotating frame
(\citealt{bt87}).  $E_{\rm J}$ is a conserved quantity even when the system is
not axisymmetric provided the system is viewed in the frame in which
the potential is stationary.  For a bar or a spiral perturbation, the
stationary frame corresponds to the one where \om{\rm p} is the pattern
speed of the perturbation.  Conservation of $E_{\rm J}$ requires that
changes in angular momentum, $\Delta L_z$, and in energy, $\Delta E$,
are related as $\Delta E = \om{\rm p} \Delta L_z$.  For most stars $\Delta
L_z$ averages to zero over an orbit; the exception is for stars at
resonances\index{resonances} (\citealt{lynden-bell_kalnajs72}).  The
change in angular momentum is accompanied by a change in the radial
action\index{radial action}, $J_{\rm R}$; in the epicyclic approximation,
this change is given by
\begin{equation}
  \Delta J_{\rm R} = \frac{\om{\rm p}-\om{}}{\kappa} \Delta L_z,
  \label{eqn:jr}
\end{equation}
where \om{}\ and $\kappa$ are the angular and radial frequencies of a
star.  \citet{sellwood_binney02} realised that changes in energy and
angular momentum at the corotation resonance\index{corotation
  resonance} (CR), where $\om{\rm p} = \om{}$, are not accompanied by any
changes in radial action, and result in no radial heating.

\begin{figure}[]
\centering{  
\includegraphics[angle=0.,width=0.7\textwidth]{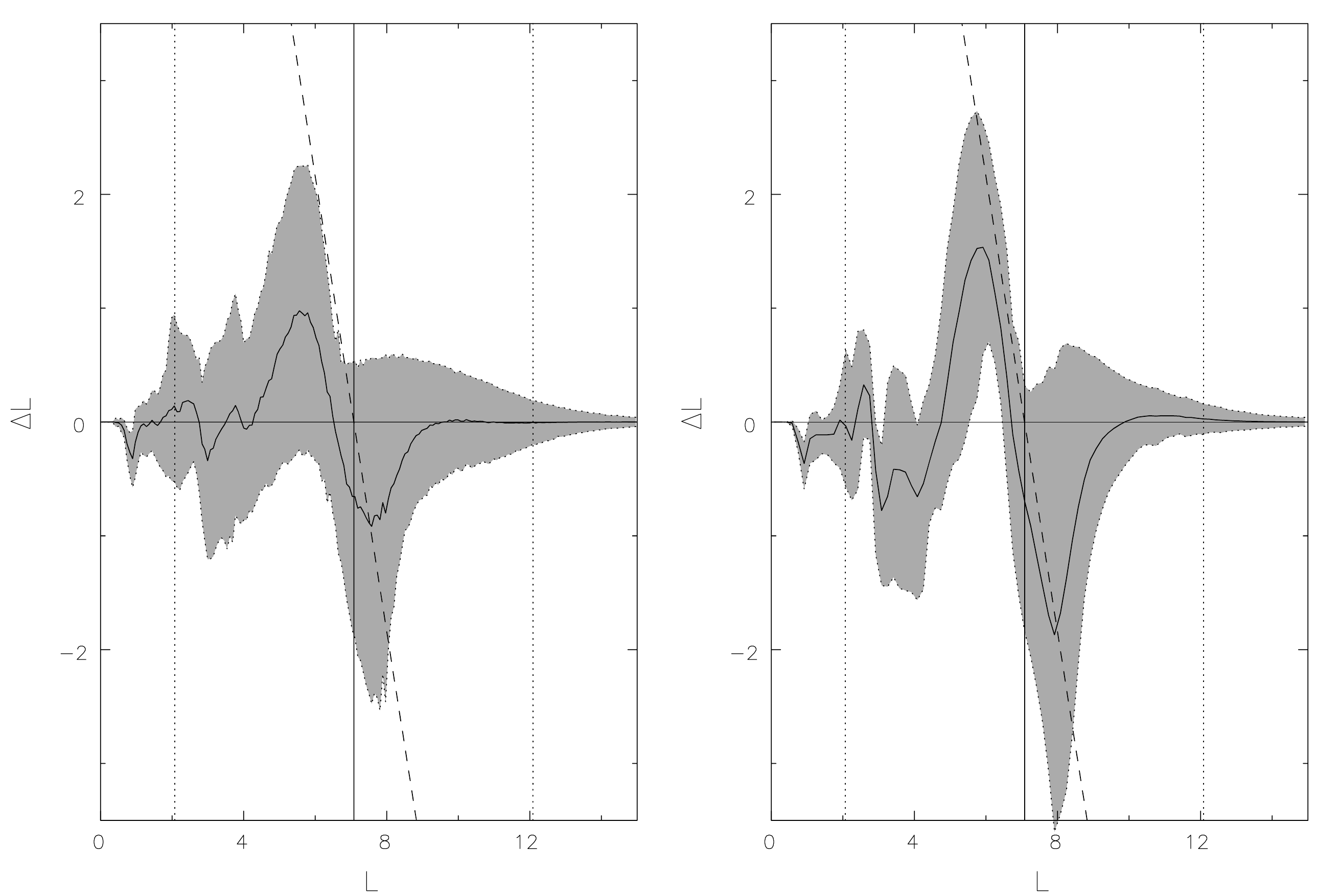}
\includegraphics[angle=0.,width=0.72\textwidth]{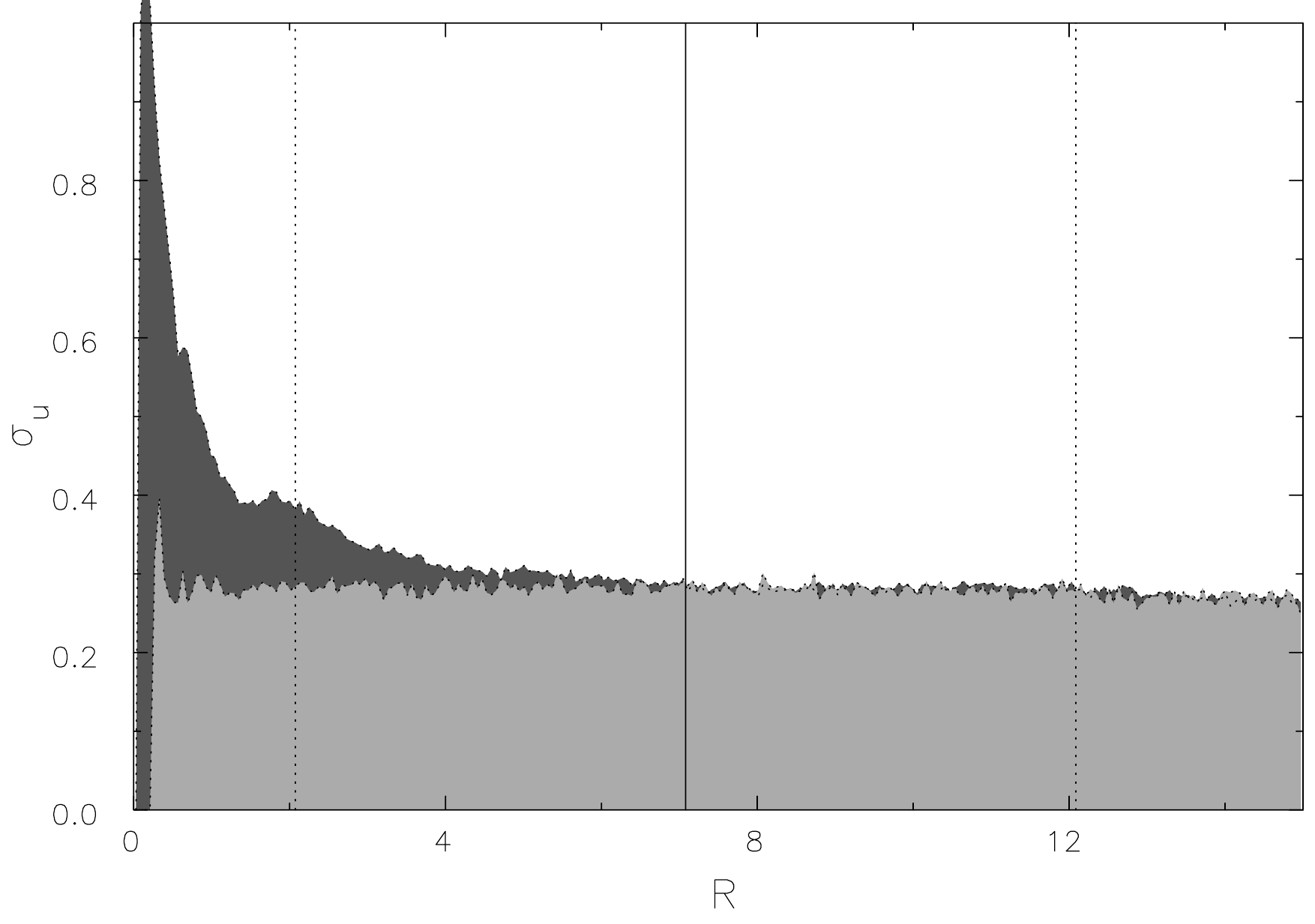}
}
\caption{{\it Top}: Exchanges of angular momentum driven by a single
  spiral.  The left panel shows angular momentum exchanges for all
  orbits while the right panel shows exchanges for nearly circular
  orbits.  The solid vertical line indicates the angular momentum of a
  circular orbit corotating with the spiral, while the dotted vertical
  lines show the ILR and OLR.  Strong angular momentum exchanges occur
  at the CR.  The shaded region shows the $20\%$ to $80\%$ interval at
  each angular momentum, with the solid line showing the mean change
  in angular momentum.
  {\it Bottom}: The radial velocity dispersion before (light grey) and
  after (dark grey) the lifetime of the spiral.  The CR (solid line),
  and the ILR and OLR (dotted lines) are indicated.  Although the
  spiral has driven substantial migration at CR, as shown in the left
  panels, the radial heating there is negligible.  Reproduced with
  permission from \citet{sellwood_binney02}}
\label{fig:sb02a} 
\end{figure}

Stars trapped\index{trapping} by a spiral at its CR change their
energy and angular momentum at fixed $E_{\rm J}$ without changing $J_{\rm R}$.
Spirals being transient (\citealt{sellwood_carlberg84, sellwood11}),
trapping by a spiral is short-lived, typically lasting for one
rotation period.  When the spiral decays, trapped stars are released
with a different angular momentum than they started with.  This is the
basis of migration due to corotation-trapping by transient spirals,
which \citet{sellwood_binney02} refer to as ``churning''\index{churning}.
Figure~\ref{fig:sb02a} presents an example of churning in a disk with a
single strong spiral.  The vertical solid lines in the top panels
show the location of the CR.  In spite of the substantial amount of
angular momentum exchanges taking place around the CR, the radial
velocity dispersion, shown in the bottom panel, at the CR is barely
altered after the spiral has died down.

\begin{figure}[]
\centering{  
\includegraphics[width=0.85\textwidth]{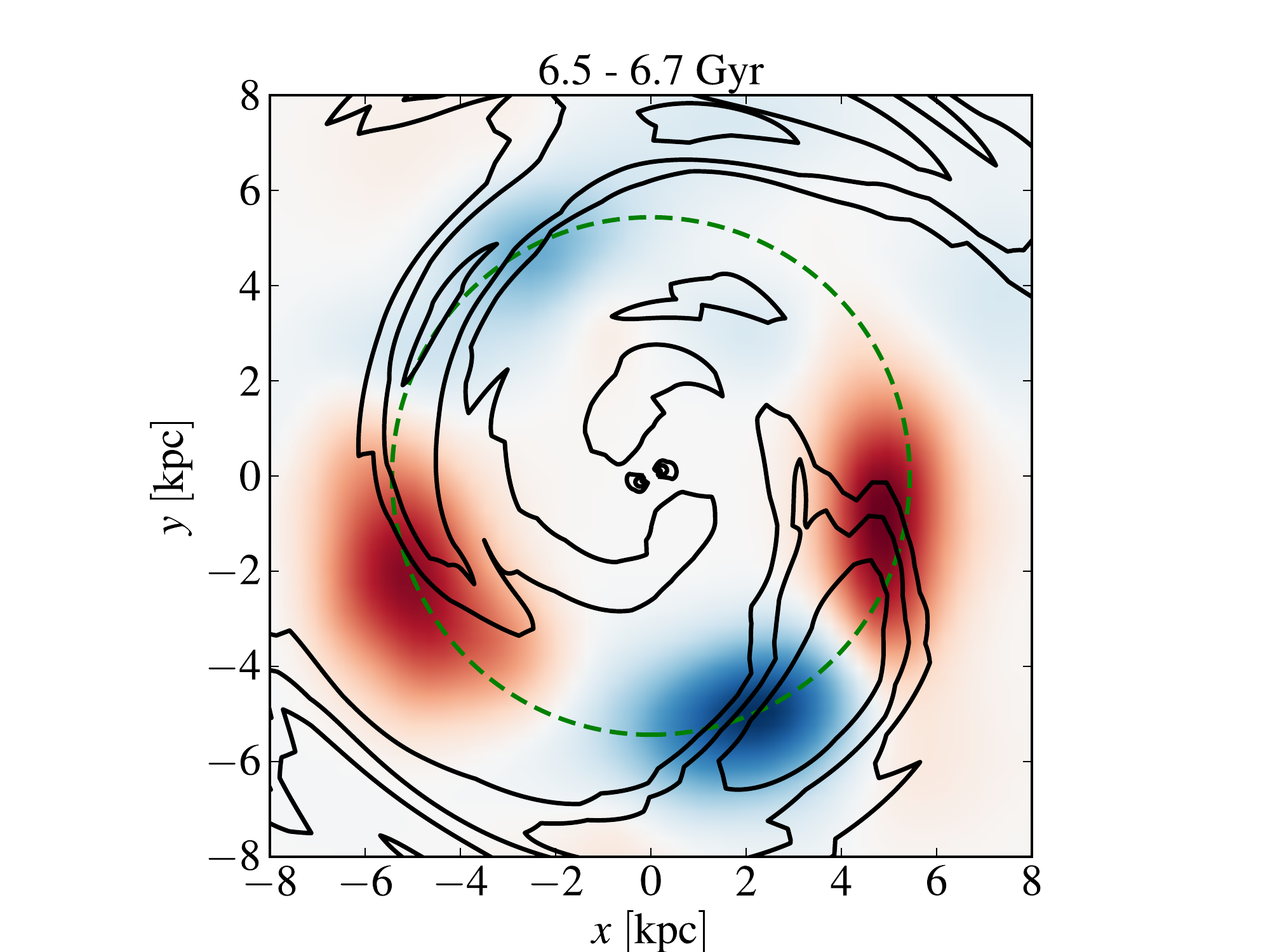}

\includegraphics[width=0.8\textwidth]{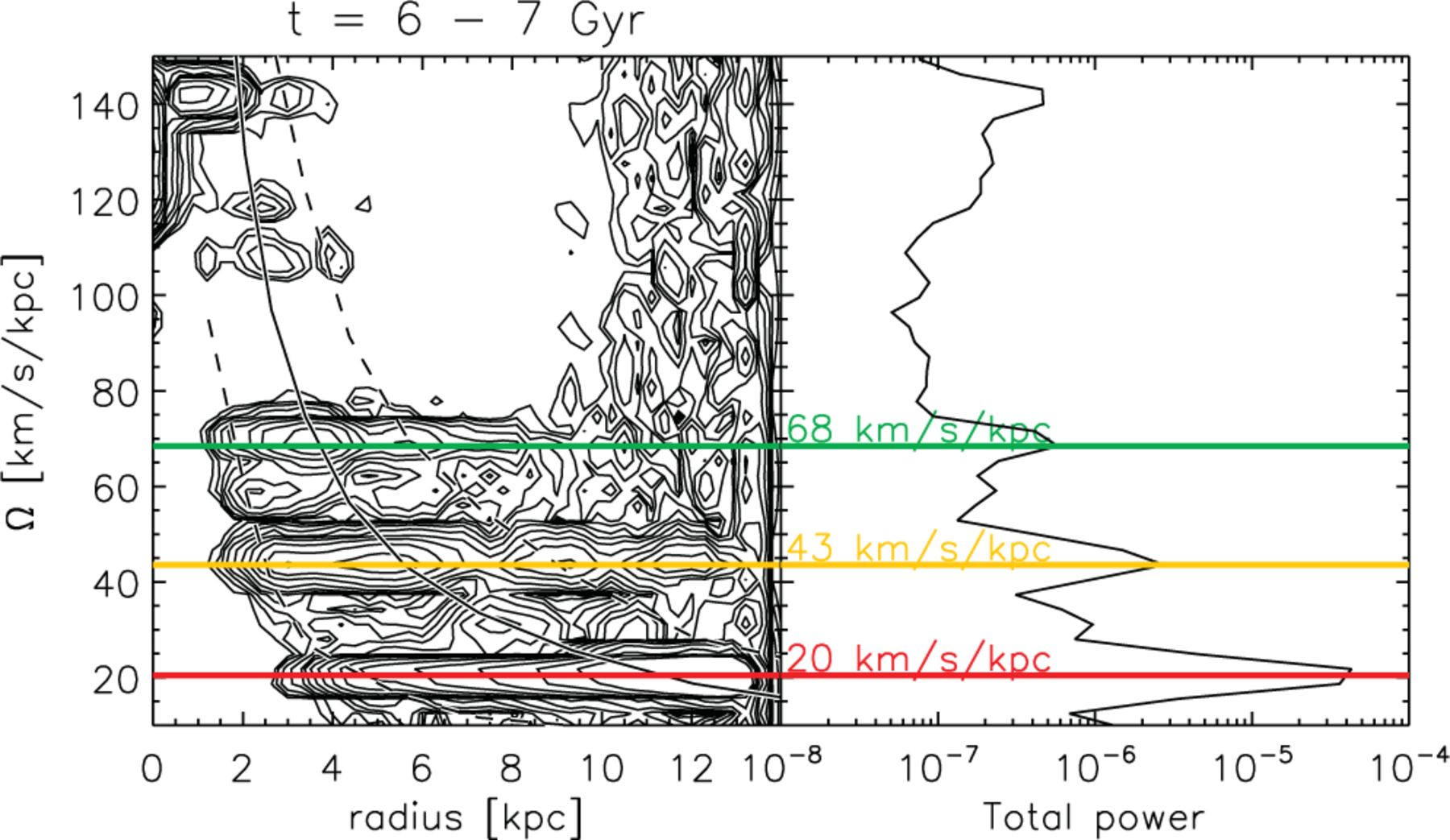}
}
\caption{{\it Top}: The location of stars relative to a growing
  spiral shortly before it reaches peak amplitude, trapping stars
  and migrating them.  The red (blue) shading indicates the density of
  stars that lose (gain) angular momentum while the contours show the
  overall density distribution.
  {\it Bottom}: The spectrum of spiral structure during the time
  interval $6-7 \,\Gyr$ for the same simulation.  Three prominent
  spirals can be seen (with pattern speeds indicated by the horizontal
  coloured lines) as well as a number of weaker ones. Reproduced with
  permission from \citet{roskar+12} }
\label{fig:roskar12a} 
\end{figure}

\begin{figure}[]
\sidecaption
\centerline{  
\includegraphics[width=0.5\hsize]{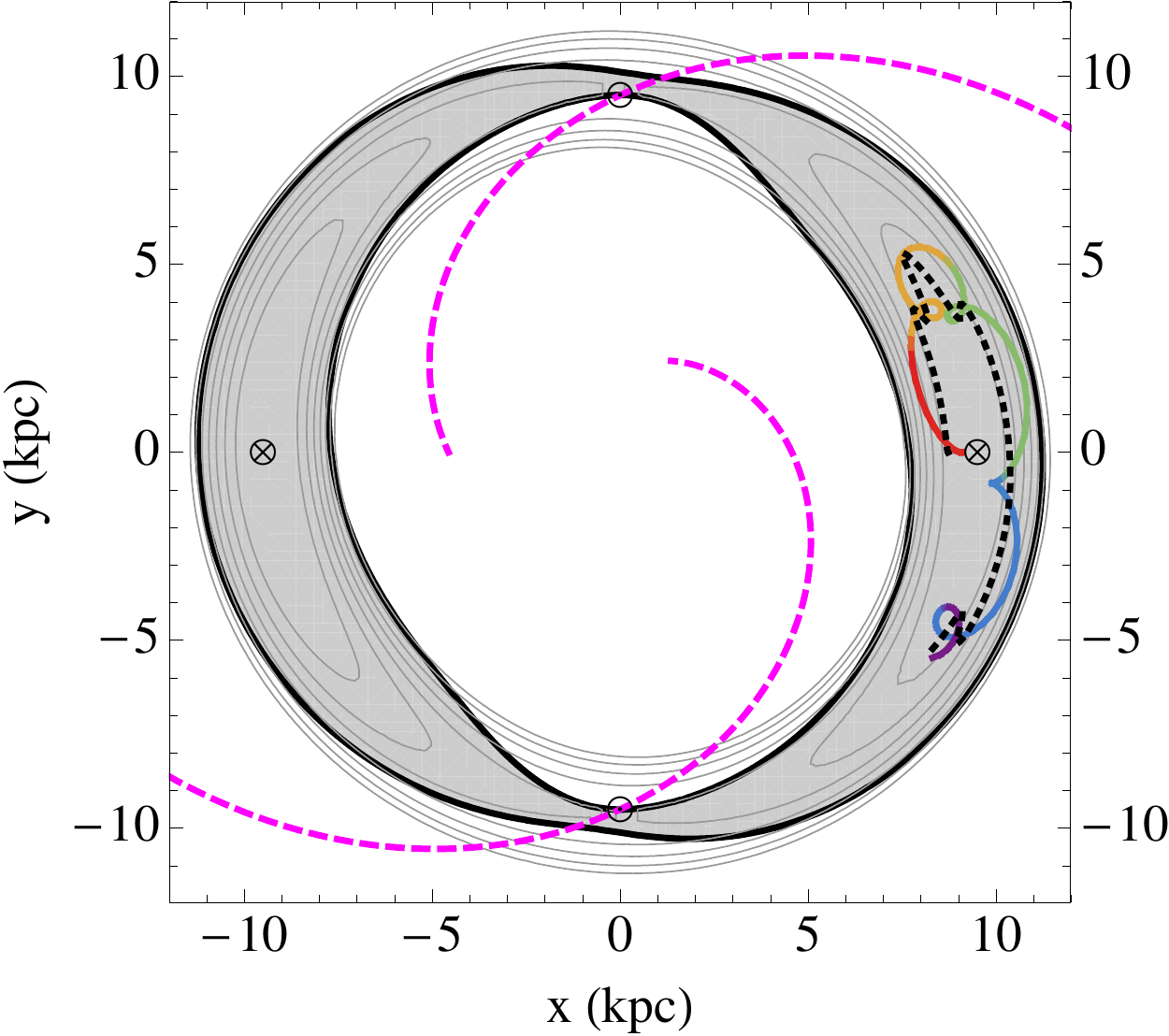}
}
\caption{Effective potential (grey contours) for a trailing $m=2$
  spiral, with peak perturbation indicated by dashed magenta lines,
  having pitch angle $\theta=25 \degrees$ and CR at $10 \,\kpc$.  The
  potential is described in \citet{daniel_wyse15}.  The crossed circles
  mark the peak of the effective potential.  The capture region is
  shaded grey.  The rainbow path is the trajectory of a trapped star
  with initial phase-space coordinates $(x,y,v_x,v_y)=(9.1 \,\kpc, 0, 0,
  -10 \,\kms)$ in the rotating frame.  The black (dotted) curve is the
  star's guiding radius.  Figure courtesy of K.J. Daniel}
\label{fig:daniel} 
\end{figure}

A simple way to understand this behaviour is to consider what happens
as a spiral density wave forms, increasing the local density in some
region.  Stars that have $\Omega > \Omega_{\rm p}$ will gain energy as they
catch up with it, but then lose it again when they overtake it.  The
net change in energy for such stars is minimal.  Conversely the stars
with $\Omega < \Omega_{\rm p}$ first lose energy, but then gain it again as
the spiral overtakes them, also with little net change in energy.
Stars that have $\Omega \simeq \Omega_{\rm p}$ never overtake, or are
overtaken by, the spiral before it decays.  These particles are
trapped by the spiral, exchanging energy and angular momentum with it.
Once the spiral subsides, their net change in angular momentum depends
on the relative phase at which they escape.  The top panel of
Fig.~\ref{fig:roskar12a} shows the location of stars that are about to
be trapped and migrated by a growing spiral.  Stars that will gain
angular momentum are located behind the spiral's peak density, while
those that will lose angular momentum are ahead of the spiral.  The
angular phase of the peak density demarcates a separation between the
gainers and the losers.  Stars are not trapped at just the corotation
radius but up to $\sim 1 \,\kpc$ from it.  \citet{daniel_wyse15} showed
that the trapping region (example shown in Fig.~\ref{fig:daniel}) is
broadened and not limited to a single radius.

\begin{figure}[]
\centerline{  
\includegraphics[angle=0.,width=0.5\hsize]{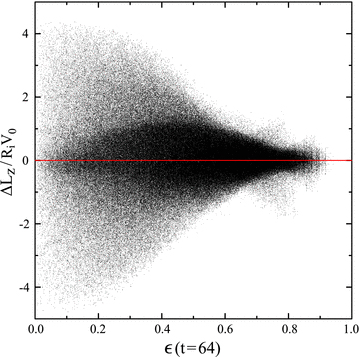}
\includegraphics[angle=0.,width=0.5\hsize]{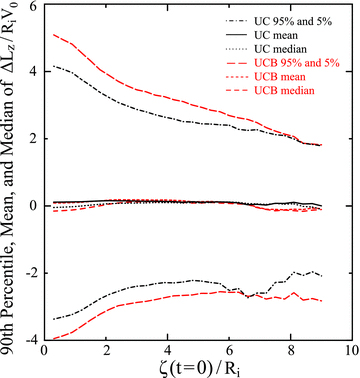}
}
\caption{{\it Left}: The effect of orbital eccentricity on migration
  efficiency.  On the horizontal axis is plotted the initial
  eccentricity of individual stellar orbits, while on the vertical
  axis is shown the normalised change in angular momentum of the
  stars.  Migration is most efficient for stars on nearly circular
  orbits.
  {\it Right}: The effect of a bar on migration.  On the horizontal
  axis $\zeta$ is a proxy for the vertical amplitude of the initial
  orbits.  The mean, median and $5\%-95\%$ of the distributions of
  angular momentum changes are shown by different lines as indicated
  at top right.  Simulation UC (black lines) did not form a bar while
  simulation UCB, with the same physical model (with $Q \simeq 1.5$)
  but different random initialisation, formed a bar.  The difference
  is computed over $\sim 6 \,\Gyr$, during which time a bar formed in
  UCB within the first Gyr.  The effect of the bar on the net
  migration is small compared to that driven by spirals.  Both
  simulations also show that the effect of height on migration is
  relatively weak.
  Reproduced with permission from \citet{solway+12}  }
\label{fig:solway12} 
\end{figure}

The efficiency of trapping\index{trapping efficiency} declines rapidly
with radial random motion (\citealt{sellwood_binney02, daniel_wyse15,
  solway+12}), but the vertical random motion reduces the trapping
efficiency much more slowly (\citealt{solway+12}).  The left panel of
Fig.~\ref{fig:solway12} plots the angular momentum change of stars as
a function of orbital eccentricity\index{orbital eccentricity},
showing that angular momentum changes become smaller as the orbital
eccentricity increases.  The right panel shows that migration is not
strongly affected by the vertical amplitude of orbits.

\subsection{Multiple Patterns}

\subsubsection{Multiple Spirals\index{multiple spirals}}

Typically disks support multiple spirals with different pattern
speeds.  The unconstrained simulation of \citet{sellwood_binney02}
supported many separate spiral frequencies, which resulted in
migration at locations across the disk.  Because the difference
between initial and final radii of stars was much larger than the
epicyclic radius, particularly at large radii, they concluded that
migration by churning was driving this migration.
\begin{figure}[]
  \sidecaption
\centerline{  
\includegraphics[width=0.5\hsize]{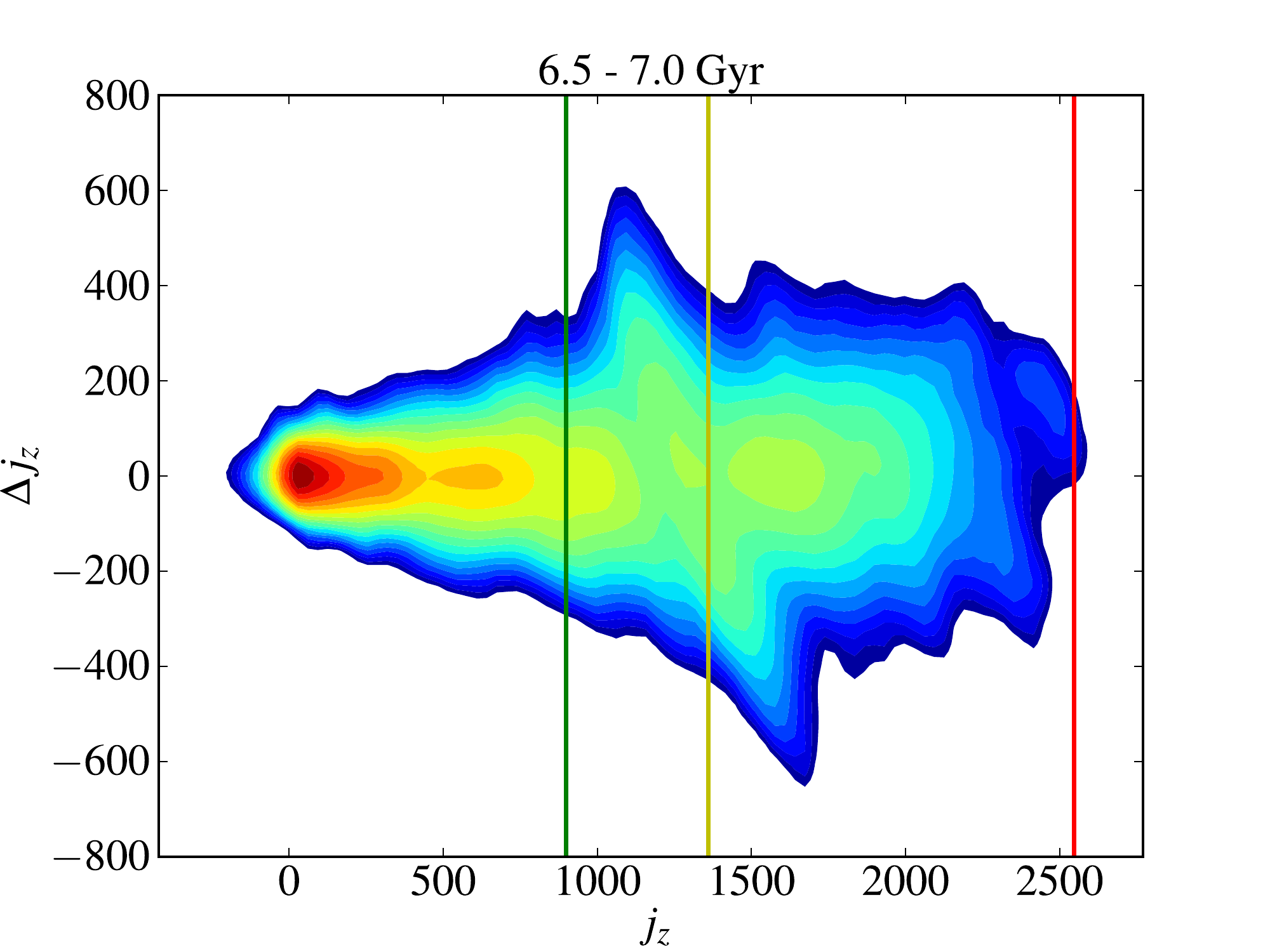}
\includegraphics[width=0.5\hsize]{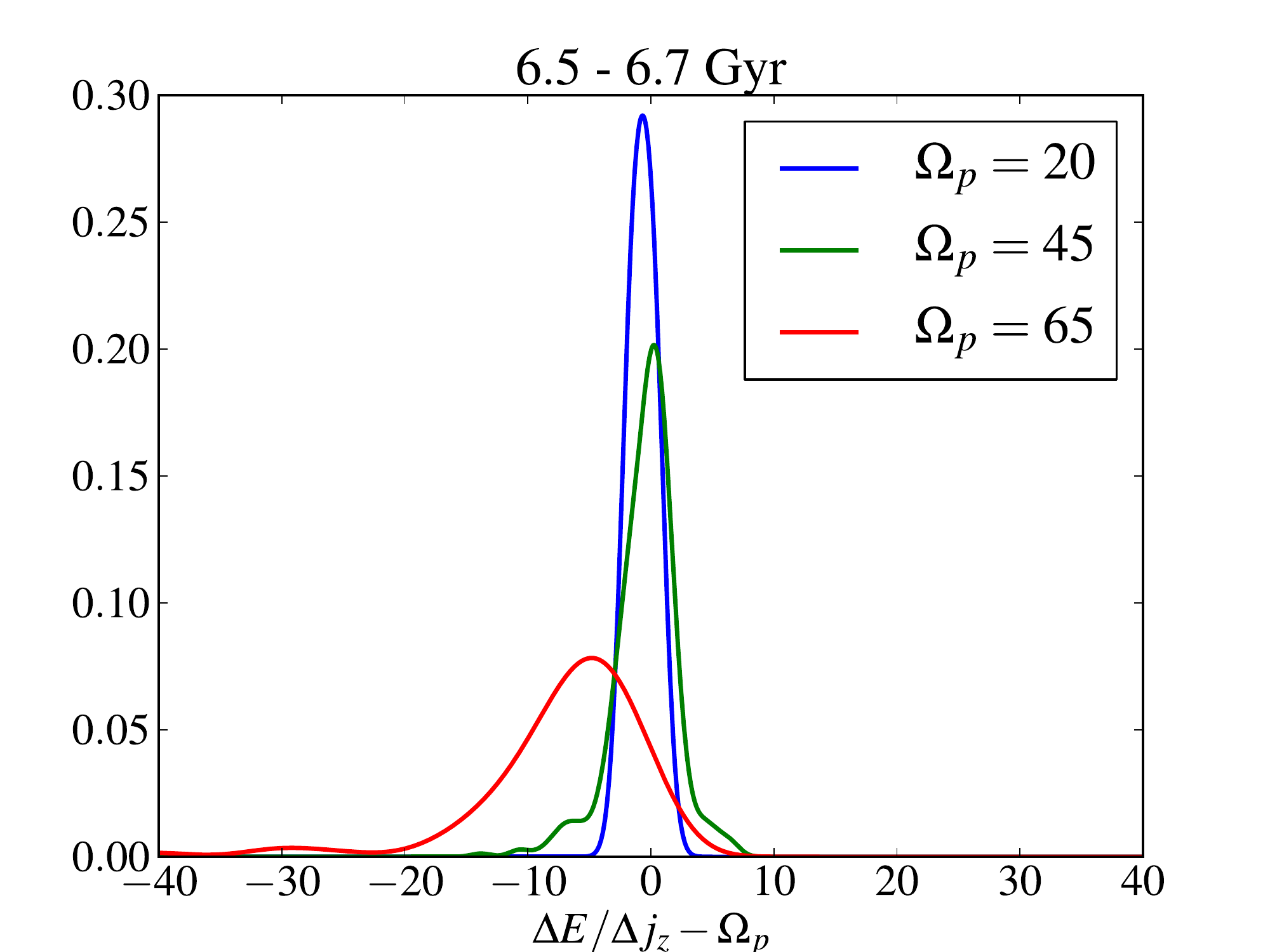}
}
\caption{{\it Left}: Distribution of changes in angular momentum as a
  function of starting angular momentum at around the same time as in
  Fig.~\ref{fig:roskar12a}.  The vertical dashed lines mark the
  locations of the CR of some of the prominent spirals.
  {\it Right}: For stars exchanging angular momentum around each of
  the main spiral CRs, the distribution of $\Delta E/\Delta j_z$ is
  strongly peaked around \om{\rm p}.  Typical uncertainties on measuring
  the pattern speeds are $\sim 5 \,\kms\,\kpc^{-1}$. Thus the Jacobi
  energy is a conserved quantity, rather than evolving chaotically,
  despite the presence of multiple patterns.  This implies that large
  angular momentum exchanges are dominated by the spiral corotation
  trapping (churning).
  Reproduced with permission from \citet{roskar+12}
}
\label{fig:roskar12b} 
\end{figure}
The simulations of \citet{roskar+12} also contained multiple spirals.
The bottom panel of Fig.~\ref{fig:roskar12a} shows a typical example:
at least three separate strong spirals with different pattern speeds
are clearly recognisable at this particular time.  The left panel of
Fig.~\ref{fig:roskar12b} shows the angular momentum exchanges
occurring during this time.  The presence of multiple spirals
increases the number of locations at which stars are migrating
(compare with the upper left panel of Fig.~\ref{fig:sb02a}).  For a
small, random sample of stars that had large migrations,
\citet{roskar+12} showed explicitly that their orbits are trapped by
the spiral at the CR.

\subsubsection{Corotating Spirals\index{corotating spirals}}

In their simulations, \citet{grand+12a} found multiple short spirals
such that most of them locally corotate with the stars at a large
range of radii.  Similar spiral behaviour was also reported by
\citet{baba+13} and \citet{roca-fabrega+13}.  \citet{sellwood_carlberg14}
demonstrated that this behaviour is a result of several high
multiplicity spirals residing in relatively low-mass disks.
\citet{grand+12a} showed that these spirals trapped a large fraction of
the stars at CR, leading to migration across a large extent of the
disk.  The extreme migrators in their simulations retain nearly
circular orbits, demonstrating that the mechanism is still that of
churning.

\subsubsection{Bars and Chaotic Scattering\index{chaotic scattering}}

The presence of multiple patterns also introduces new physics.
\citet{daniel_wyse15} showed that a star trapped at CR can escape if
its guiding radius approaches an inner or an outer Lindblad resonance
(ILR or OLR) of another pattern, without the spiral amplitude
changing.  This changes the rate of migration; unlike trapping at the
CR, however, this is a scattering process, raising the random motion
of stars.

Enhanced migration in the presence of multiple patterns, but
especially when a bar is present, has been proposed by a number of
studies.  \citet{minchev_famaey10} suggested that coupling between
multiple patterns\index{coupling between multiple patterns} (the
combination of a bar and spirals, or between multiple spirals) drives
strong, chaotic, migration which is substantially more efficient than
that driven by churning.  The simulations of \citet{minchev+11}
exhibited such rapid migration that stellar populations were
substantially mixed on a $3 \,\Gyr$ timescale.  Contrary to
\citet{debattista+06} and \citet{foyle+08}, they suggested that
bar-driven migration is persistent.  \citet{brunetti+11} also found
fast migration (on timescales of order a rotation period) in the case
of a bar forming in a cool ($Q \sim 1$) disk, but mixing was much
reduced if the initial disk is hot (see also \citealt{debattista+06}).
\citet{halle+15} studied radial redistribution using a bar-unstable
simulation, distinguishing between the effects of heating and of
churning on the basis of the mean radius of particle orbits.  They
argued that the bar's corotation is the main source of migration.  In
the outer disk they found heating dominates, rather than migration, by
a factor of $\sim 2-8$.

Other studies have questioned whether such strong migration occurs as
a result of multiple patterns.  A quantitative assessment of the
impact of bars on migration was obtained by \citet{solway+12}.  The
right panel of Fig.~\ref{fig:solway12} compares the degree of
angular momentum changes in two models, which represent the same
physical initial conditions but, because of the stochasticity inherent
in bar formation (\citealt{sellwood_debattista09}), in one a strong bar
formed while in the other no bar formed.  The bar formed during the
first \Gyr; after $\sim 6\,\Gyr$ of evolution, the change in
angular momentum is larger when a bar is present, but not by the order
of magnitude predicted by \citet{minchev+11}.

Nor is there evidence for strong scattering resulting from the
presence of multiple spirals.  \citet{roskar+12} tested explicitly for
chaotic migration in their simulations with multiple spiral arms (see
Fig.~\ref{fig:roskar12a}).  They found smoothly evolving orbital
frequencies, a characteristic of regular, not chaotic, evolution.
Moreover they showed that the changes in energy and angular momentum
conserve the Jacobi energy in the frame of the spirals, as shown in
the right panel of Fig.~\ref{fig:roskar12b}.  $E_{\rm J}$ would not be
conserved if the chaotic interaction of two or more patterns was
scattering the stars, since no stationary closed frame then exists.

The observational evidence also does not favour extremely efficient
mixing driven by bars.  For instance \citet{sanchez-blazquez+14}
measured stellar age and metallicity gradients in a sample of barred
and unbarred galaxies.  They found no significant difference in their
gradients, arguing that migration by bars is not strong enough to
erase gradients.

\subsubsection{Churning versus Scattering}

The word ``migration'' has come to mean the motion of stars from one
radius to another either because of trapping at corotation by spirals
(churning) or by a variety of heating mechanisms.  As
\citet{sellwood_binney02} noted, the measured velocity dispersion of
old stars allows very little room for random motions in the Solar
neighbourhood to have moved stars by more than $\Delta R \sim 1.3
\,\kpc$.

{\it
The key difference between churning and scattering is in the
random motion created, with churning not increasing random motions
appreciably while scattering leads to disks becoming hotter.
}

\subsection{Evidence for Migration in the Milky Way}

Because we can study it in much greater detail than other galaxies,
thus far the evidence for migration has been strongest in the Milky
Way\index{Milky Way}:

\begin{itemize}
\item The age-metallicity relation\index{Milky Way: age-metallicity
  relation} of the Milky Way is flat and broad (\citealt{haywood+13,
  bergemann+14, rebassa-mansergas+16}), contrary to the expected
  evolution from traditional chemical evolution models.
  \citet{sellwood_binney02} argued that these are characteristics of
  migration, as has been shown by models (e.g., \citealt{roskar+08b,
    schoenrich_binney09a}).

\item The radial \feh\ gradient\index{Milky Way: metallicity gradient}
  decreases with stellar age (\citealt{yu+12}).  Since the metallicity
  gradient is expected to be steeper at earlier times, this flattening
  must result from migration smearing the metallicity profile of older
  stars (\citealt{roskar+08b, schoenrich_binney09a}).

\item \citet{bovy+16} showed that low-\alfe, mono-abundance
  populations\index{Milky Way:mono-abundance populations} flare
  radially, which is the expected behaviour for populations that
  migrate while conserving the vertical action (\citealt{solway+12,
    vera-ciro_donghia16}).
  
\item The skewness of the metallicity distribution
  function\index{Milky Way: MDF skewness} varies across the disk,
  changing from skewed to low \feh\ in the inner disk to skewed to
  high \feh\ at larger radii (\citealt{hayden+15}).  This pattern arises
  because at larger radii stellar populations include an increasing
  fraction of metal-rich stars that have migrated outwards from small
  radii (\citealt{hayden+15, loebman+16}).  
  
\end{itemize}

This partial list demonstrates that a broad range of observations
favour migration having occurred in the Milky Way.  Whether this is
because of churning or scattering is still not decided.  The
realisation that metal-rich stars, which are probably outward
migrators, become proportionally more common at radii past the Solar
cylinder hints that the outer disk is a repository of migrated stars.


\section{Type~II Profiles}
\label{sec:typeII}

The type~II profile is the most common amongst star-forming galaxies.
Two obvious potential causes for a steeper exponential profile in the
outer disk are obscuration by dust and a slowly declining star
formation rate (SFR). In an H$\alpha$ spectroscopic study of 15
edge-on galaxies, \citet{christlein+10} excluded both of these
possibilities.  Their sample was free of surface brightness
enhancements and the data clearly showed a cutoff in the SFR much
steeper than that of the stellar continuum. Therefore another
mechanism is required to explain the mass in the outer disks.

\citet{pohlen_trujillo06} and \citet{erwin+08} classified type~II
profiles into various classes based on the break radius and the origin
which this suggests.  They distinguished between inner and outer
breaks, the latter separated into breaks arising from asymmetric
disks, classical breaks at large radius and, in barred galaxies, ones
at roughly twice the bar radius, which they associated with the bar's
OLR.  \citet{pohlen_trujillo06} found classical breaks in $40\%$ of Sc
and later-type galaxies.

\subsection{Theoretical Models}

Thus type~II breaks are probably due to a number of mechanisms; here
we concentrate on classical breaks which are associated with star
formation breaks\index{star formation breaks}.  In the standard
paradigm of galaxy formation, galaxies grow by accreting gas and
forming stars (\citealt{fall_efstathiou80, mo_mao_white}).  Because gas
accreted at later times has higher angular momentum, it lands at
larger radii (\citealt{guo+11}).  Galaxies should therefore form stars at
increasingly large radii, which is termed ``inside-out
formation''\index{inside-out formation}.  Inside-out formation predicts
that young stars should be present in the outskirts of galaxies
(\citealt{larson76, matteucci_francois89, chiappini+97, naab_ostriker06}),
as is observed inside the break (\citealt{dejong96IV, bell_dejong00,
  macarthur+04, munoz-mateos+07}).

\subsubsection{Star Formation Breaks}

\begin{figure}[]
\centerline{  
\includegraphics[angle=0.,width=0.54\hsize]{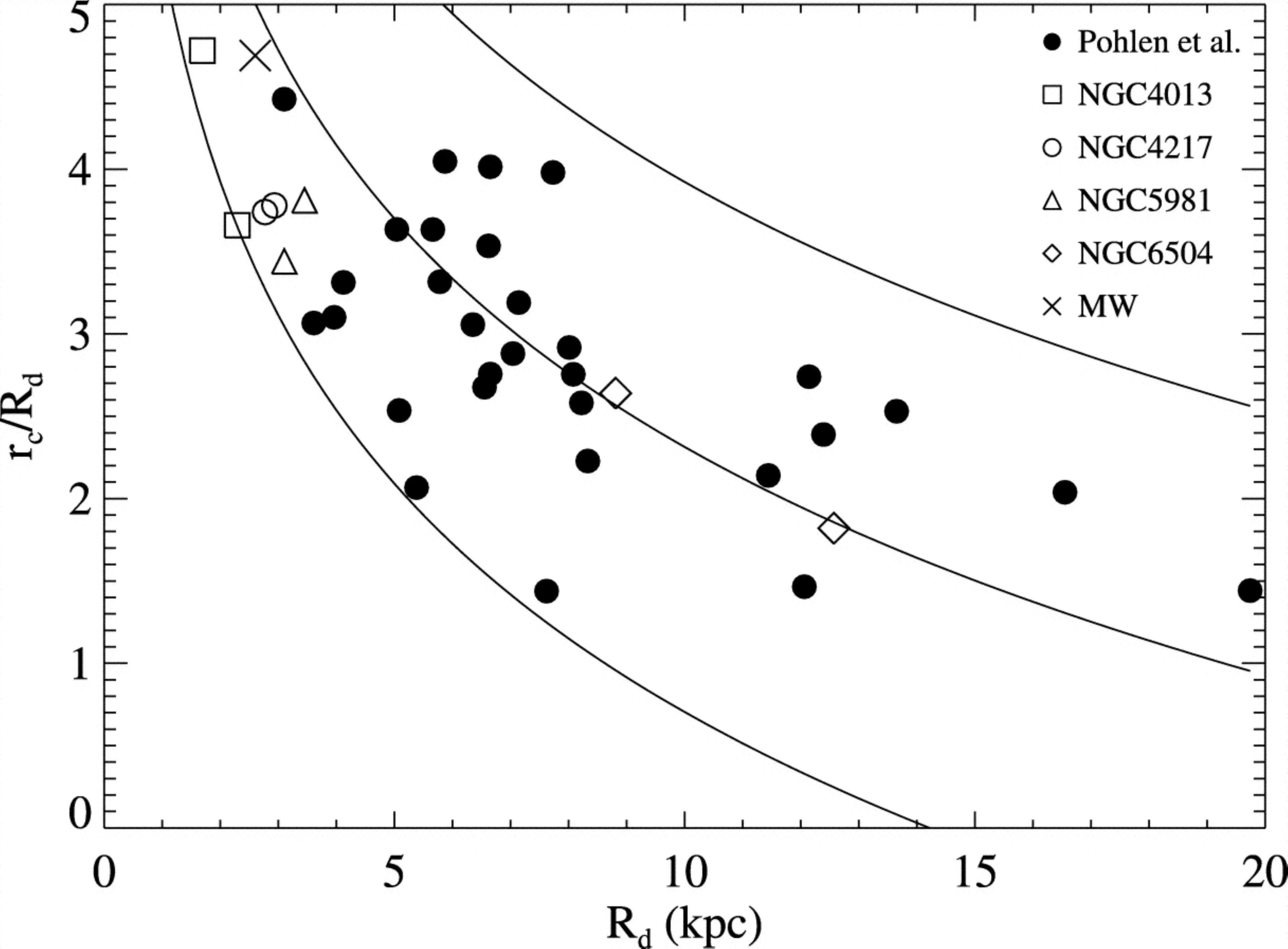}
\includegraphics[angle=0.,width=0.46\hsize]{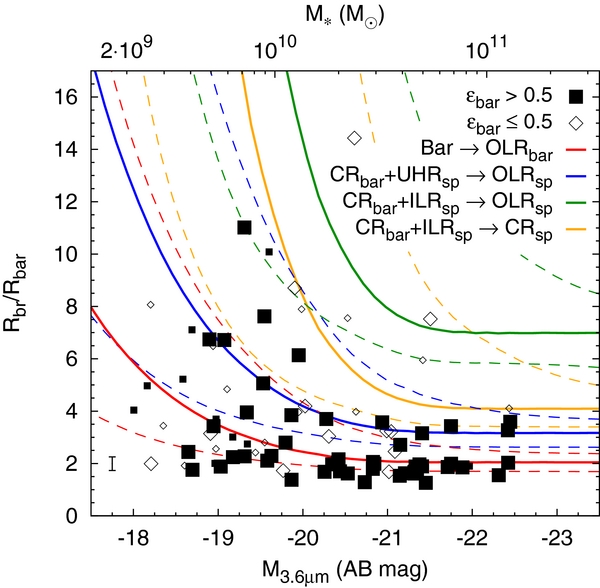}
}
\caption{{\it Left}: Break radius, $r_{\rm c}$, as a function of \Rd, in the
  model of \citet{schaye04}.  The solid lines correspond to disks of
  different masses ($M_\mathrm{disk} = 7.5 \times 10^9\,\Msun$, $3.8
  \times 10^{10}\,\Msun$, and $1.9 \times 10^{11}\,\Msun$, from top to
  bottom) assuming a surface density of the phase transition at $\log
  (N_\mathrm{H,crit}\,\mathrm{cm}^2) = 20.75$.  Observational datapoints
  from the literature are indicated.  However, both sides are well fit
  by models at fixed galaxy mass.  Reproduced with permission from
  \citet{schaye04}.
  {\it Right}: Evidence for the effect of bars on the locations of
  disk breaks.  Filled squares and open diamonds correspond to strong
  and weak bars.  The red solid curve shows the locus of the bar's OLR
  while the blue solid line corresponds to the OLR of a spiral which
  has its inner 4:1 resonance at the bar's CR.  The yellow and green
  lines are the OLR and CR of spirals with ILR at the bar's CR.
  Reproduced with permission from \citet{munoz-mateos+13}}
\label{fig:schaye04} 
\end{figure}

Star formation in disk galaxies typically drops abruptly at a few \Rd,
even though the atomic gas extends beyond this point
(e.g., \citealt{kennicutt89, martin_kennicutt01, bigiel+10}).  Efforts
to associate stellar breaks with star formation breaks started early
(e.g., \citealt{vdkruit87, ferguson+98}).  \citet{schaye04} proposed that
disk breaks occur as a result of star formation thresholds caused by
the phase transition of gas from warm ($T \sim 10^4$~K) to cold ($T
\sim 10^2$~K).  Once this thermal instability sets in, it is quickly
followed by a gravitational instability due to the associated sharp
drop in Toomre-$Q$, allowing star formation to proceed.
\citet{schaye04} computed where the stellar breaks should occur, and
found a good match to observed stellar break radii, as shown in
Fig.~\ref{fig:schaye04}.  Thus breaks in the stellar distribution show
a strong relation to breaks in the present day star formation.

The model of \citet{schaye04} assumed that the disk is axisymmetric.
Recognising that any perturbations that arise in the outer disk will
enhance the local gas density, he predicted that the thermal
instability will occur, triggering star formation in the outer disk.
Thus it is not surprising to find star formation at large radii,
beyond $2\,R_{25}$ (e.g., \citealt{gildepaz+05, thilker+05,
  zaritsky_christlein07, christlein_zaritsky08}), with extended UV
(XUV)\index{XUV disks} emission in about $20\%$ of galaxies
(\citealt{lemonias+11}), although it is unlikely that all XUV disks can be
explained this way.
\citet{schaye04} also neglected the effects of bars, which are
efficient at modifying the distribution of gas.  Bars accumulate gas
at the OLR (e.g., \citealt{schwarz81, byrd+94, rautiainen_salo00}).  A
bar may also influence the gas further out via resonances of spirals
coupled to the bar\index{bar-spiral coupling}.  Both these effects are
present in the barred sample of \citet{munoz-mateos+13}, who showed
that breaks tend to follow two loci (shown in the right panel of Fig.
\ref{fig:schaye04}): one given by the OLR of the bar, and the other by
the OLR of a spiral that has its 4:1 resonance at the CR of the bar.
In both cases the effect of bars is to alter the gas distribution and
thereby the location of star formation breaks.
Finally, star formation need not truncate sharply:
\citet{elmegreen_hunter06} proposed a multicomponent star formation
model with star formation in the outer disk predominantly driven by
turbulent processes (see the review by Elmegreen \& Hunter, this volume.)

\subsubsection{Angular Momentum Redistribution Models\index{angular momentum redistribution}}

\begin{figure}[]
\centerline{  
\includegraphics[angle=0.,width=0.43\hsize]{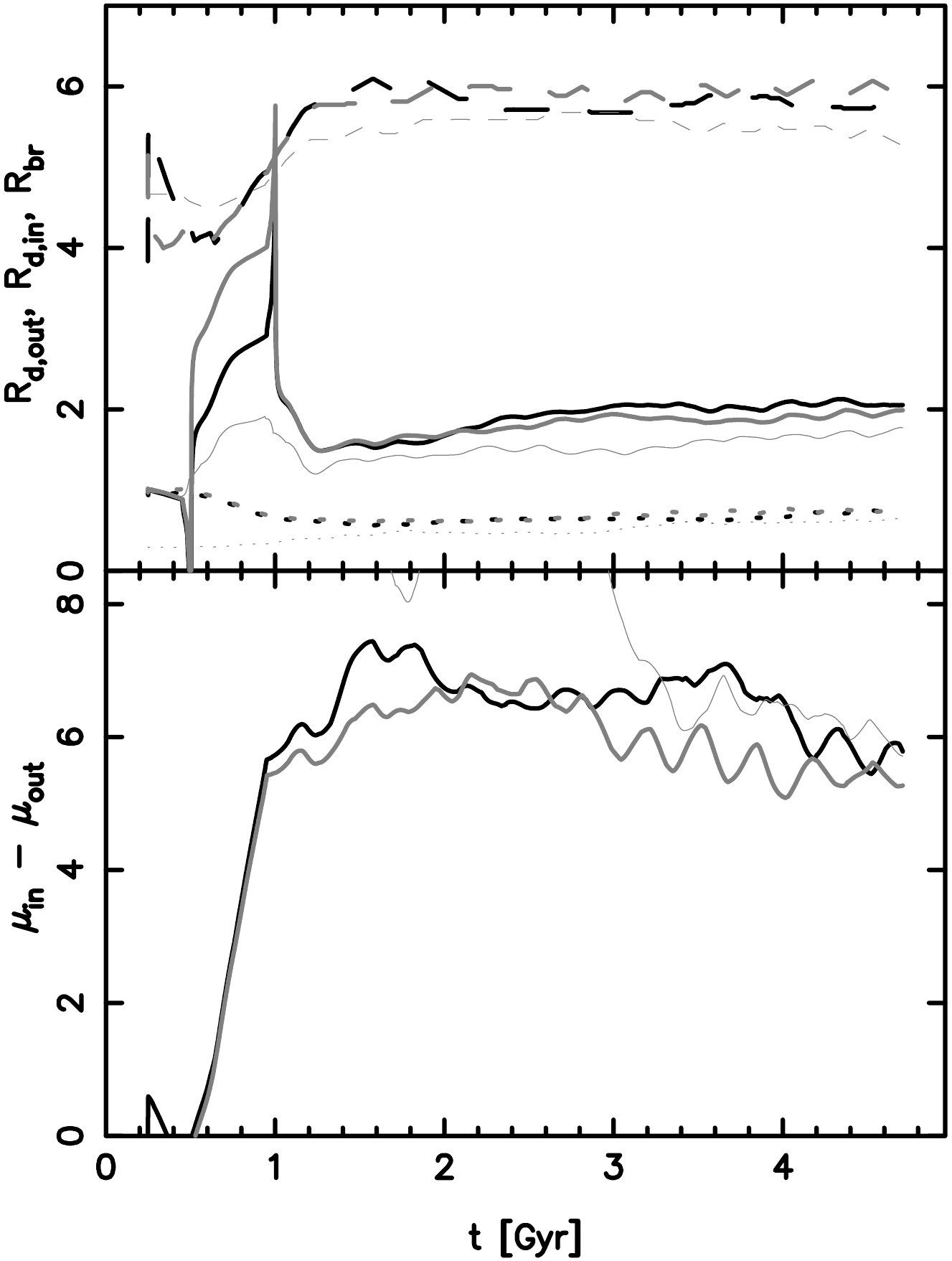}
\includegraphics[angle=0.,width=0.57\hsize]{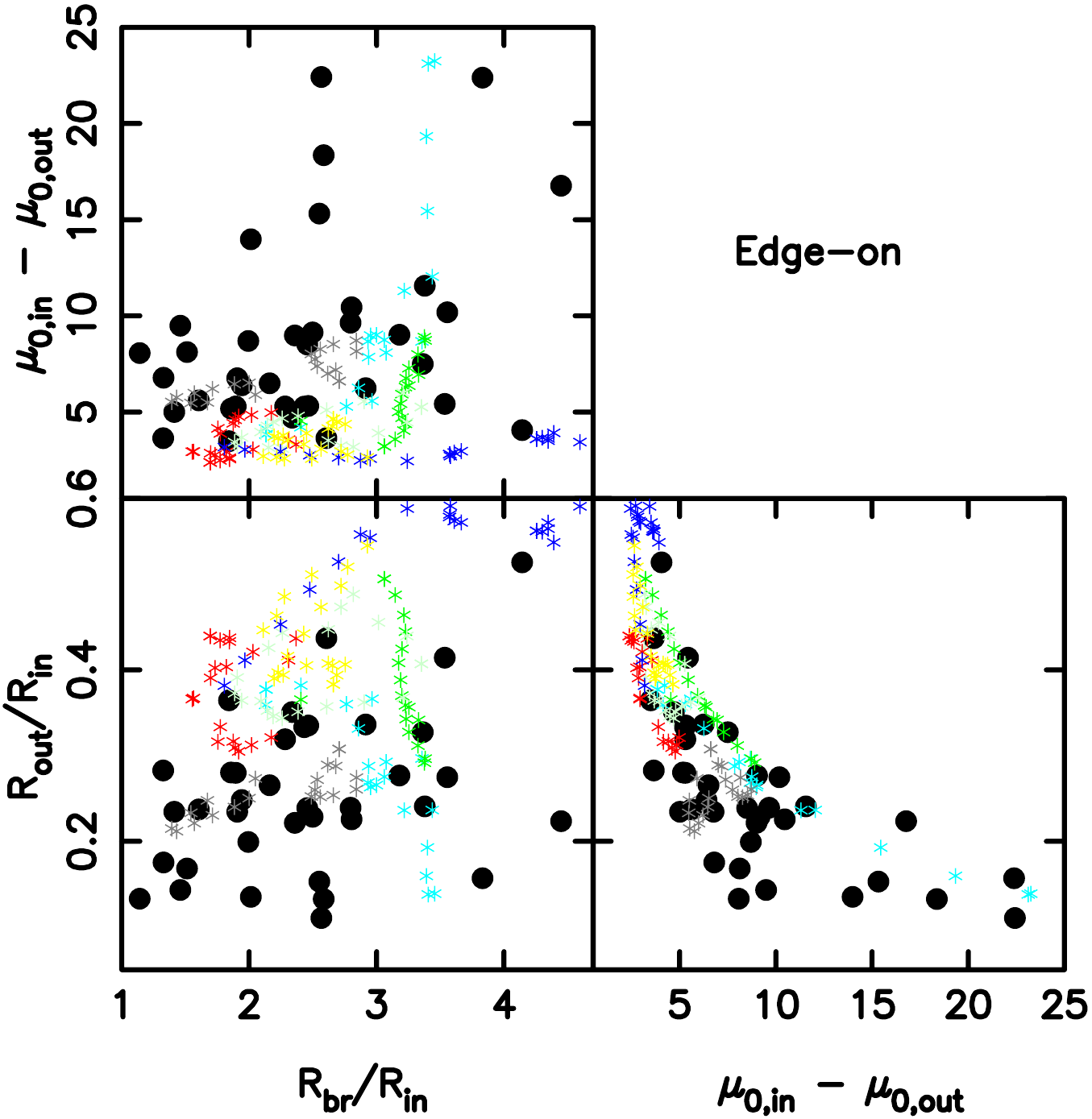}
}
\caption{{\it Left}: The evolution of break parameters in the pure
  $N$-body simulations of \citet{debattista+06}: $R_{\rm br}$ (the radius
  of the break; dashed lines), $R_{\rm d,in}$ and $R_{\rm d,out}$ (the scale
  lengths of the inner and outer disks; solid and dotted lines,
  respectively), and the difference in central surface brightness
  between the inner and outer disks, $\mu_{\rm in} - \mu_{\rm out}$.  In these
  simulations the bar forms by $1 \,\Gyr$.  The black, grey and thin
  lines correspond to different {\it initial} radial extents of the
  same model.
  {\it Right}: A comparison between parameters of profile breaks in
  observed (solid circles) and simulated (coloured stars) edge-on
  galaxies.  The observational data are taken from \citet{pohlen_phd}.
  Each simulation corresponds to a different colour and different
  viewing angles are shown by different points.  Reproduced with
  permission from \citet{debattista+06}  }
\label{fig:debattista06} 
\end{figure}

Since most stars do not form in the outer disks, stars may reach there
from the inner disk.  The first mechanism that was proposed to explain
type~II profiles sought to transform type~I profiles via angular
momentum redistribution due to coupled bars and spirals
(\citealt{debattista+06}).  Their pure $N$-body simulations produced
spirals with inner 4:1 resonance coinciding with the CR of the bar;
angular momentum was transported to the OLR of the spiral, where a
break in the profile developed.  This angular momentum redistribution
occurs {\it during bar formation}, and decreases thereafter.  The left
panel of Fig.~\ref{fig:debattista06} shows that the break barely
evolves after the first $1.5 \,\Gyr$, during which time the bar forms.
The structural parameters of the resulting breaks are in good
agreement with observations, as shown in the right panel of
Fig.~\ref{fig:debattista06}.  \citet{debattista+06} found that the
inner disk needs to be cool ($Q \ltsim 1.6$) in order for bar
formation to require enough angular momentum redistribution to
significantly alter the density profile, whereas in hotter disks
strong bars formed without a significant break developing.  The idea
was developed further by \citet{foyle+08}, who conducted a large
simulation study of bulge$+$disk systems, including gas and star
formation.  They found that when breaks form, the outer profile remains
unchanged while the inner disk changes substantially.  They reported
that type~II profiles developed when $m_{\rm d}/\lambda \ge 1$, where $m_{\rm d}$
is the disk mass fraction and $\lambda$ is the halo angular momentum.
The inclusion of gas led to a break in star formation, but this did
not always result in a break in the total profile.  In agreement with
\citet{debattista+06}, they found that the location of the break does
not evolve much once it formed.  In a similar spirit,
\citet{minchev+11} found that, within $1-3 \,\Gyr$, as bars form, the
density profile of their models extended outwards to more than
$10\,\Rd$, with a flattened metallicity distribution.  \citet{minchev+12}
emphasize that outer disks forming via this bar-spiral resonance
overlap are radially hot.

The formation of type~II profiles by angular momentum redistribution
during bar formation presents a number of difficulties; most obviously
strong bars are not present in all galaxies.  \citet{minchev+11} argued
that bars are a recurrent phenomenon and that the absence of a bar now
does not mean that one was never present.  However, bars are not easily
destroyed (e.g., \citealt{shen_sellwood04}).  Additionally bar
destruction\index{bar destruction} leaves a kinematically hot disk; a
re-formed bar would subsequently lead to reduced angular momentum
transport (\citealt{debattista+06, brunetti+11}).  The formation of
type~II profiles in such simulations is likely significantly affected
by the very unstable initial conditions usually employed, which may be
unlikely in nature.

Another way of scattering material into the outer disk was proposed by
\citet{bournaud+07}, who used simulations of gas rich disks which form
clumps\index{clumps} similar to those observed in chain galaxies.
These clumps interact gravitationally with each other, flinging
material outwards while sinking to form a central bulge.  The material
flung out forms a kinematically hot outer exponential of a type~II
profile.  The existence of bulgeless galaxies with type~II profiles
(e.g., NGC~3589, NGC~6155, UGC~12709; \citealt{pohlen_trujillo06}) is a
challenge for this model to explain type~II profiles observed at low
redshift since clump formation inevitably also leads to bulge
formation.

\subsubsection{Churning\index{churning}}

Alternatively, \citet{roskar+08a} proposed that the outer disk in
type~II profiles results from stars that have migrated past the star
formation break via churning.  Stellar density breaks formed in this
way are coincident with a drop in the star formation rate.  In their
simulations, stars formed from gas continuously cooling from the
corona rather than being introduced into the simulation as part of the
initial conditions.  Because of this, the disk heats as it is forming,
remaining at $Q \simeq 2$ through most of its extent and never becomes
violently unstable, while still being cool enough to continuously
support multiple spirals.  \citet{roskar+08a} found that $\sim 85\%$ of
the stars that end up beyond the break formed interior to it, yet are
on nearly circular orbits, with epicyclic radii half of the average
radial excursion, which is because spiral trapping is most efficient
for nearly-circular orbits.

\begin{figure}[]
\centering{  
\includegraphics[angle=0.,width=0.7\textwidth]{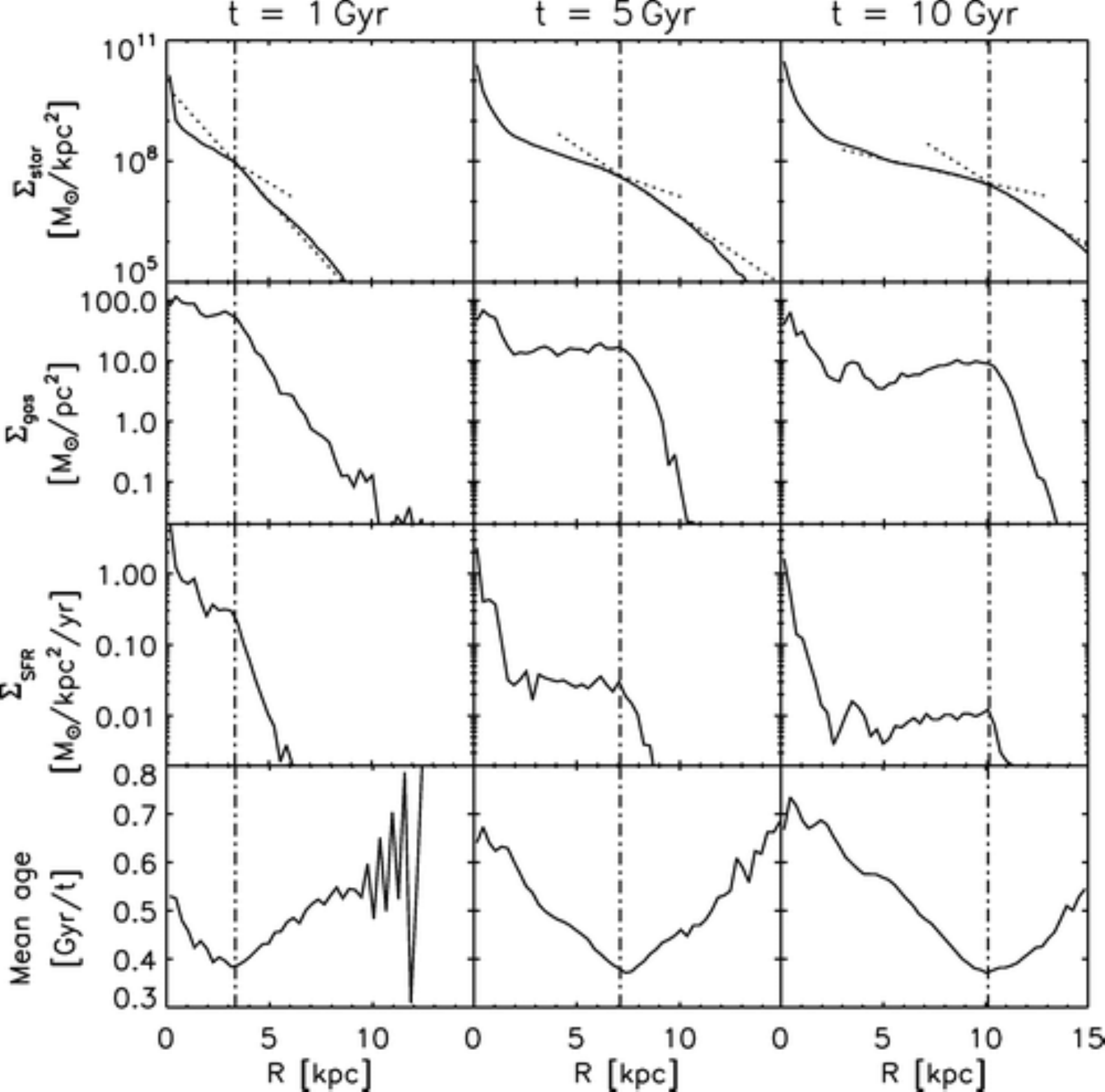}
}
\caption{Evolution of the system with stars forming from
  gas cooling off a hot corona.  The {\it top} row shows the stellar density
  profile, the {\it second} row shows the surface density of cool gas, the
  {\it third} row shows the star formation rate surface density and the
  {\it bottom} row shows the average age of stars, normalised to the age of
  the galaxy at that time.  From {\it left} to {\it right} the columns show the
  system after $1$, $5$, and $10~\,\Gyr$.  The vertical dot-dashed lines
  show the stellar break radius while the dotted lines in the top row
  show the inner and outer exponential disk fits
}
\label{fig:roskar08a} 
\end{figure}

\begin{figure}[]
\centering{  
\includegraphics[angle=0.,width=0.8\textwidth]{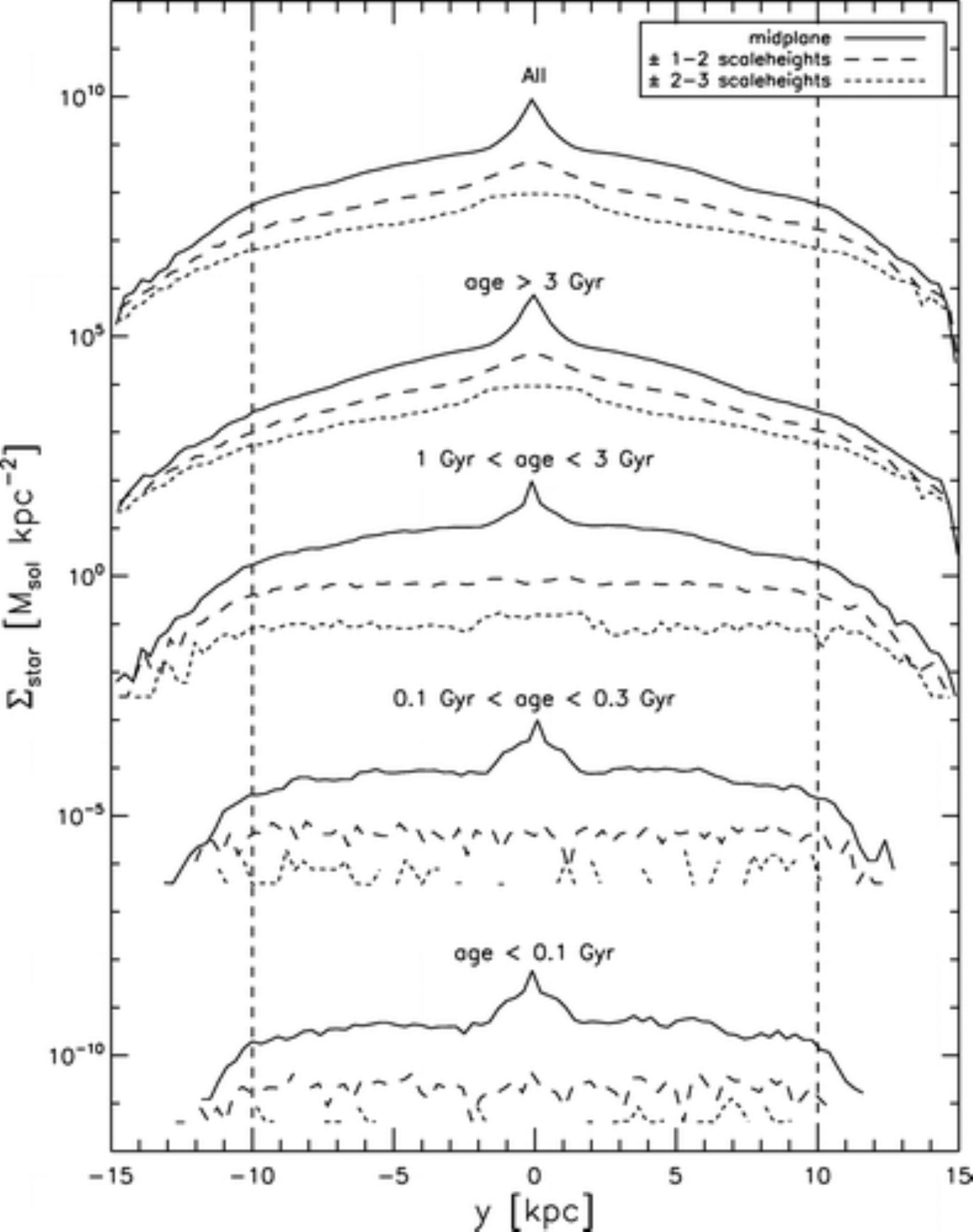}
}
\caption{The density distribution of stars for the galaxy on the
  left at $10 \,\Gyr$, observed edge-on.  Different age bins are shown,
  as indicated, with the top profiles showing the profiles of all the
  stars.  In each set of profiles, three heights above the mid-plane
  are shown, as indicated in the inset.  An arbitrary offset has been
  applied to each set of profiles, for clarity.  Reproduced with
  permission from \citet{roskar+08a}  }
\label{fig:roskar08a_bis} 
\end{figure}

{\it
In the churning model of type~II profiles, while the break radius
grows with time, churning forces the break to occur at the same radius
for stars of all ages, and all heights above the plane, as seen in Fig.~\ref{fig:roskar08a_bis}.  The model also predicts
that the mean age of stars beyond the star formation break {\it
  increases} outwards, as shown in Fig.~\ref{fig:roskar08a}.  This
age upturn\index{age upturns} occurs because churning, being
stochastic, requires increasingly long times to populate regions
further from where stars form.
}

\subsubsection{Insight from Cosmological Simulations}

\citet{sanchez-blazquez+09} studied breaks using a cosmological
simulation.  The break in the simulation was caused by a drop in the
star formation due to a warp, which led to a drop in the volume
density of the gas.  The stars beyond the break were predominantly
($57\%$) formed interior to the break, with the remainder coming
equally from in-situ formation and from satellite
accretion\index{satellite accretion}.  They interpreted the age upturn
in their simulation as resulting from a low but constant star
formation at large radii, rather than to migration.
\citet{ruiz-lara+16b} also found age upturns in their cosmological
simulations, which were caused by satellites accreted onto the outer
disks.  An observational signature of such accretion is that past the
break the age upturn is followed within $\sim 1\,\Rd$ by an age plateau.

However, cosmological simulations are still plagued by excessively hot
disks\index{hot disks}.  \citet{sanchez-blazquez+09} dealt with this by
identifying the disk component via a kinematic cut on stellar orbits
to retain stars on nearly circular orbits.  This resulted in a meagre
$37\%$ of stars being sufficiently cool to be considered as disk
stars.  Likewise \citet{ruiz-lara+16b} applied a kinematic cut to the
stellar particles, but were left with a hot system with even young
populations being as hot as the old stars in the Solar neighbourhood.
\citet{roskar+10} stressed that however the disk is defined, one is
left with a high Toomre-$Q$ system which substantially inhibits spiral
structure (e.g., \citealt{binney_tremaine08}), suppressing churning;
therefore these simulations cannot yet give a clear indication of the
relative contribution of accreted, scattered and migrated by churning
stars in outer disks.

\subsection{Observational Tests}

\subsubsection{Resolved Stellar Population\index{resolved stellar populations} Studies}
\label{sssec:resolvedpops}

\begin{figure}[h!]
\includegraphics[angle=0.,width=0.5\hsize]{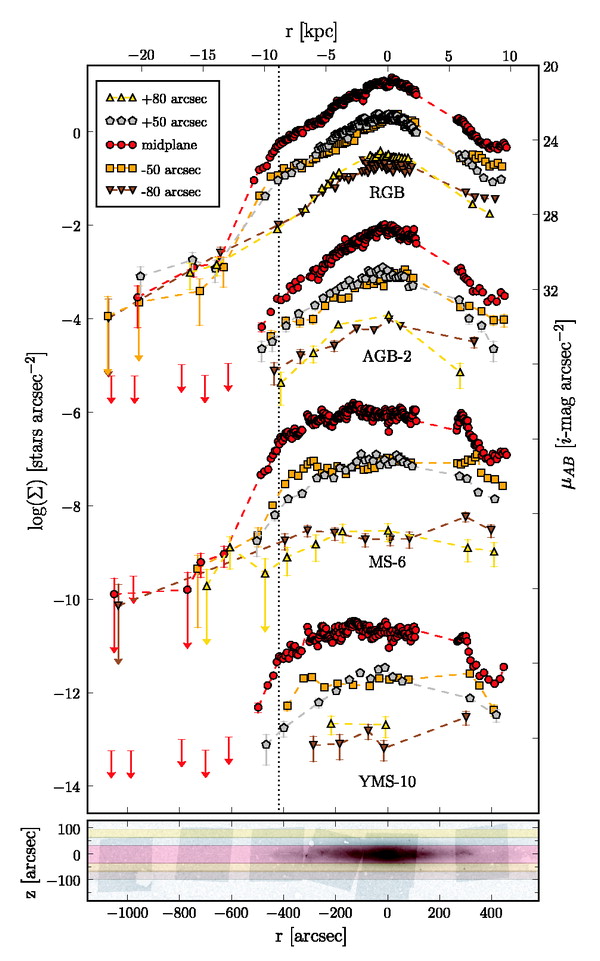}
\includegraphics[angle=0.,width=0.5\hsize]{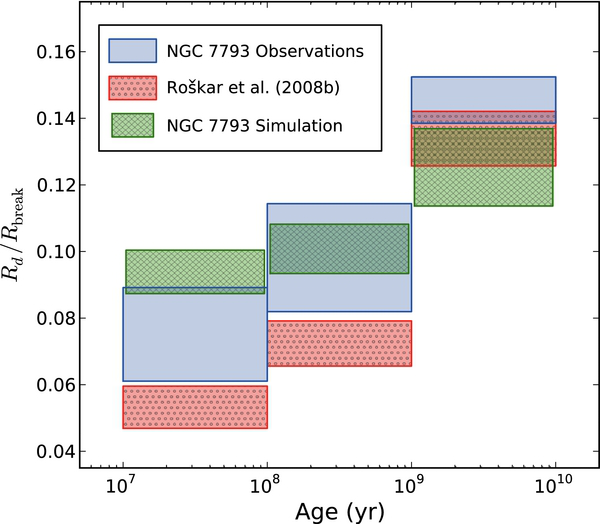}               %
\caption{{\it Left}: Radial density profiles for resolved stellar
  populations in the edge-on NGC~4244 (shown at bottom).  The
  different populations are young main sequence (YMS) stars ($< 100
  \Myr$), main sequence (MS) stars ($100-300 \Myr$), asymptotic giant
  branch (AGB) stars, (older than $0.3\,\Gyr$ and peaking at $1-3 \,\Gyr$
  with a tail to $10 \,\Gyr$) and metal poor red giant branch (RGB)
  stars (older than $5 \,\Gyr$).  Five strips are used, as indicated at
  top-left, at three different offsets from the mid-plane.  The AGB,
  MS and YMS profiles have been offset vertically by -2, -6 and -10
  respectively, for clarity.  This can be compared with
  Fig.~\ref{fig:roskar08a_bis}.  Reproduced with permission from
  \citet{dejong+07}.
  {\it Right}: The outer disk exponential scale lengths, normalised
  by the break radius, as a function of age for NGC~7793 (shaded blue
  box) and two simulations: the Milky-Way mass one of
  \citet{roskar+08b} (dotted red box) and another of the same mass and
  size as NGC~7793 (hatched green box).  The height of the boxes
  represents the uncertainty on the scale-length.  Reproduced with
  permission from \citet{radburn-smith+12}  }
\label{fig:djrs} 
\end{figure}

Resolved stellar population studies are ideal for assessing the
populations of outer disks because they can be observed directly and
little modelling is necessary.  An early study of stellar populations
across the break came from the GHOSTS survey (\citealt{radburn-smith+11})
for the edge-on galaxy NGC~4244 (\citealt{dejong+07}).  They separated
stars into four age bins at different heights above the mid-plane.
They found that stars of all ages have a break at the same radius,
regardless of whether they are in the mid-plane or above it, as seen
in Fig.~\ref{fig:djrs}.  Unless the break formed very recently (for
no apparent reason), or formed once and has not evolved since
(whereas observations show that break radii have evolved since
  $z\sim 1$, e.g., \citealt{perez04, trujillo_pohlen05, azzollini+08b}) the
most plausible explanation is that the break radius evolves with time,
with something forcing all stars to adopt a common break radius.  The
churning model predicts exactly this behaviour: as seen in
Fig.~\ref{fig:roskar08a}, the break radius in the simulations of
\citet{roskar+08a} evolves with time, yet the location of the break is
identical for all populations.

A second example of a type~II profile with a common break radius for
all ages is the low-inclination flocculent spiral galaxy NGC~7793
(\citealt{radburn-smith+12}).  Here, the break is just inside the break of
both the {\sc Hi} gas, and of the star formation.  Beyond the break
$\Rd$ increases with the age of the stellar population.
Fig.~\ref{fig:djrs} shows the scale-length of different age bins for
NGC~7793, and for a simulation of a galaxy of comparable mass and size
which experienced significant churning.

M33 also has a type~II profile with a break at $\sim 8\,\kpc$
(\citealt{ferguson+07}), which is present also in the surface {\it mass}
density (\citealt{barker+11}); stars past the break exhibit a positive age
gradient (\citealt{barker+07, williams+09a, barker+11}).  Unlike NGC~4244
and NGC~7793, M33 has had a strong interaction, in the past $1-3 \,\Gyr$
(\citealt{braun_thilker04, putman+09, davidge_puzia11}).  \citet{barker+11}
found that about half of all stars just inside the break have ages
between $2.5\,\Gyr$ and $4.5\,\Gyr$, with less than $14\%$ of stars older.
While the star formation within $8 \,\kpc$ does not appear to have
changed much in the past few \Myr, the main sequence luminosity
function of the outer disk indicates that the recent star formation
rate in the outer disk has declined (\citealt{davidge_puzia11}),
suggesting enhanced star formation in the outer disk or contamination
by young stars.  In the past $10\,\Gyr$ $\Rd$ has increased by a factor
of $\sim 2$.  \citet{bwilliams+13} contrasted M33 with its near equals
in mass, NGC~300 and NGC~2403, both relatively isolated bulgeless
late-type galaxies with type~I profiles.  They concluded that the environment
has played a role in the formation of the break in M33.

\subsubsection{Spectroscopic Studies}

\begin{figure}[]
\centerline{  
\includegraphics[angle=0.,width=0.5\hsize]{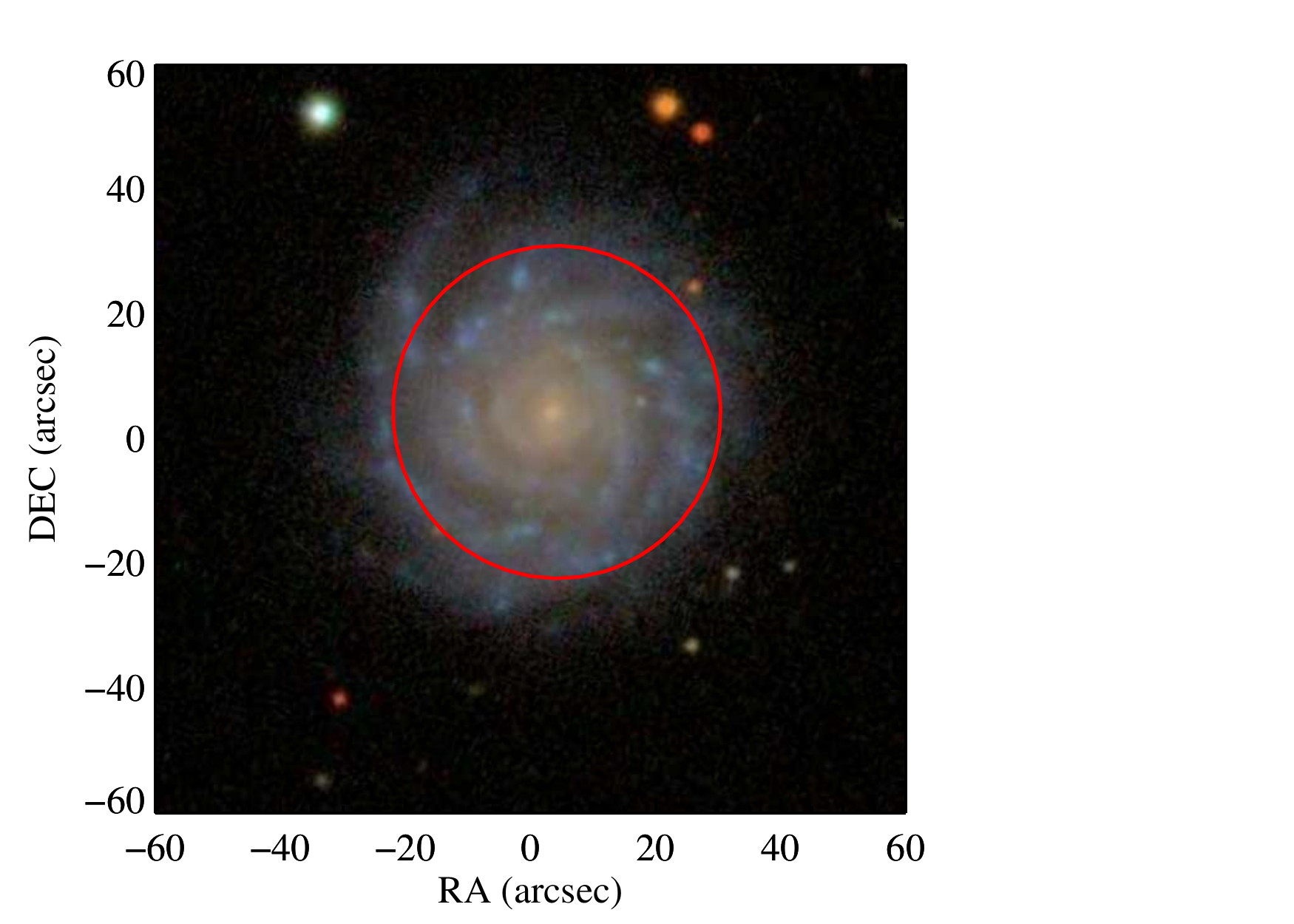}
\includegraphics[angle=0.,width=0.5\hsize]{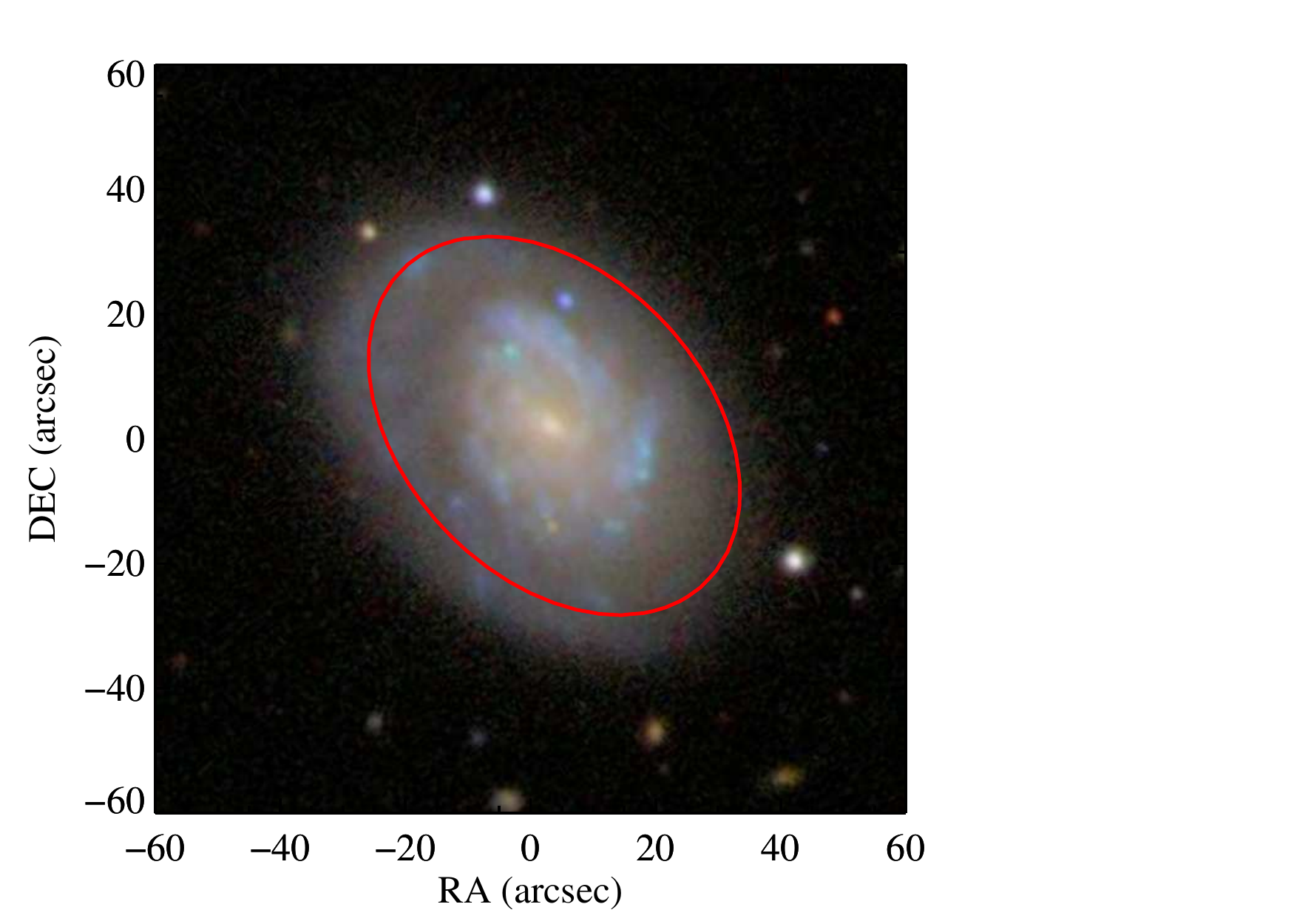}
}
\centerline{  
\includegraphics[angle=0.,width=0.5\hsize]{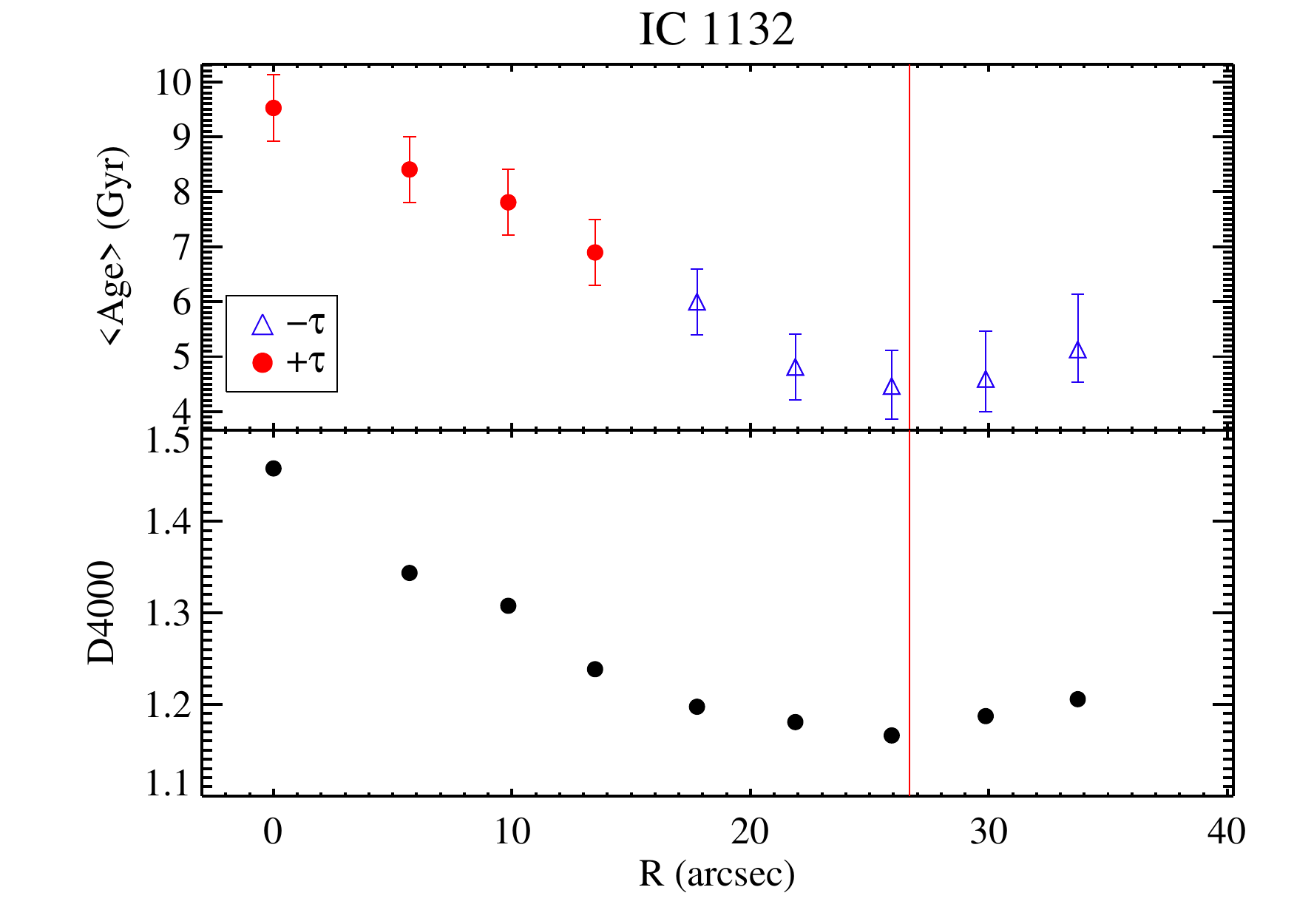}
\includegraphics[angle=0.,width=0.5\hsize]{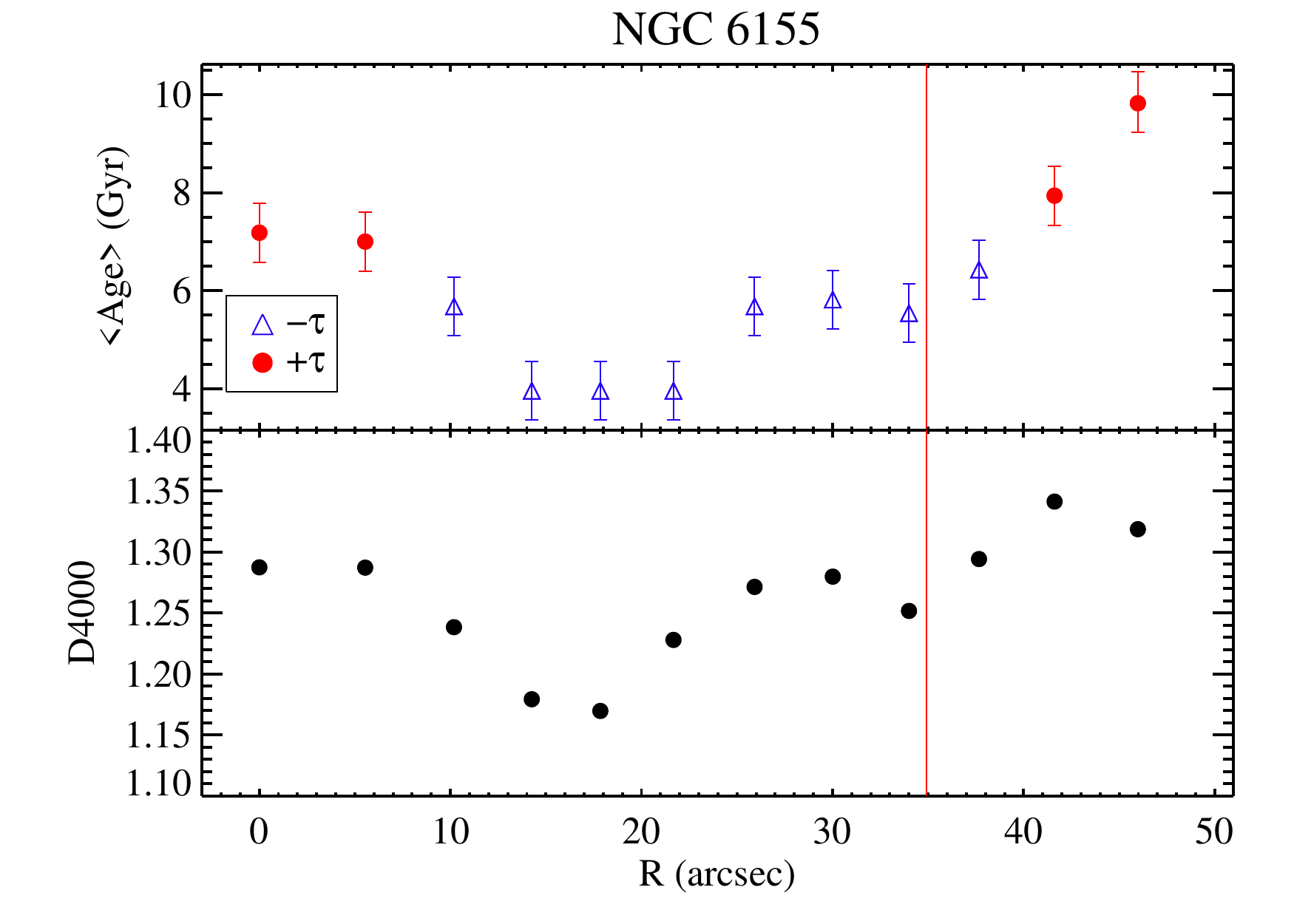}
}
\caption{{\it Left}: IC~1132.
  {\it Right}: NGC~6155.  The {\it top} row shows images of the galaxies
  with the location of the break indicated by the red ellipses.  The
  {\it bottom} row shows the mean age profiles (top panel) and the D4000
  index ({\it bottom} panel).  In the mean age profiles, blue points
  correspond to $\tau < 0$ (exponentially increasing SFR), while the
  red points correspond to $\tau > 0$ (exponentially decreasing SFR).
  Reproduced with permission from \citet{yoachim+12} }
\label{fig:yoachim12} 
\end{figure}

\citet{yoachim+12} modelled integral field data for six type~II profile
galaxies, obtaining their star formation history.  In all cases the
average age profile has a positive gradient beyond the break.  The
modelling fitted an exponentially declining SFR,
$\psi(t) \propto e^{-t/\tau}$,
where the time-scale, $\tau$, was allowed to take both positive
(declining SFR) and negative (increasing SFR) values.  In three of the
six galaxies, $\tau$ was positive, as expected if the stars in this
region migrated there.  The other three galaxies had negative $\tau$,
indicating an increasing SFR, which is not expected if the outer disk
formed exclusively from migrated stars.  Fig.~\ref{fig:yoachim12}
presents an example of each behaviour.  NGC~6155, shown on the right,
is a galaxy where the stellar populations past the break have a
radially increasing average age and, except for the innermost bin,
temporally decreasing star formation rate.  While the average stellar
age increases past the break, the minimum age is not at the break
itself, but half way between the centre and the break; inspection of
the top panel of Fig.~\ref{fig:yoachim12} shows that, at the location
of the age minimum, the galaxy has a strongly star-forming ring
encircling a bar.

{\it
  Churning predicts that the mean age increases past the break.
  However, the break need not be the location of the minimum in average
  age, which happens only if inside-out growth is exact.
}

Another example of an age minimum before the break radius is M95, a
barred galaxy with a ring at $\sim 8\,\kpc$ previously classified as having an
OLR type~II profile (\citealt{erwin+08}).  However, the star formation
break is at $\sim 13\,\kpc$, beyond which the colour gradient reverses,
extending to $\sim 25\,\kpc$ (\citealt{watkins+14}).

For three galaxies, \citet{yoachim+12} found that the star formation
timescale remains negative (increasing SFR) to the last measured
point.  The left panels of Fig.~\ref{fig:yoachim12} show an example,
IC~1132.  What appears to be happening in each of these galaxies is
that new stars are forming in-situ due to spirals extending past the
break radius, as envisaged by \citet{schaye04}.  Since the SFR does not
terminate at the break, either in the observations or in the
simulations (\citealt{roskar+08a} found that 15\% of stars formed
  in-situ), it should not be surprising to find hints of
\mbox{recent} star formation past the break.  Moreover, such galaxies are
easier to measure spectroscopically so may be over-represented in a
small sample.

\citet{ruiz-lara+16} studied the stellar populations of 44 face-on
spiral galaxies from the CALIFA survey.  They found that $\sim40\%$ of
these show age upturns beyond the break in the luminosity-weighted
spectra.  They found very flat mass-weighted mean age profiles, with
less than 1 Gyr variation in NGC~551 and NGC~4711.  Moreover, they
found that luminosity-weighted mean age upturns are present in both
type~II and type~I profiles.  \citet{roediger+12} had also seen age
upturns in type~I profiles for galaxies in the Virgo
Cluster\index{Virgo Cluster}.  This suggests that age upturns in
type~I galaxies may be a result of environment\index{environment} (see
Sect.~\ref{sec:typeI} below).  Their presence in type~I profiles led
\citet{ruiz-lara+16} to propose that age upturns are the result of the
early formation of the entire disk followed by a gradual inside-out
quenching of star formation.

\subsubsection{Photometric Methods}

\begin{figure}[h!]
  \sidecaption
\includegraphics[angle=0.,width=\hsize]{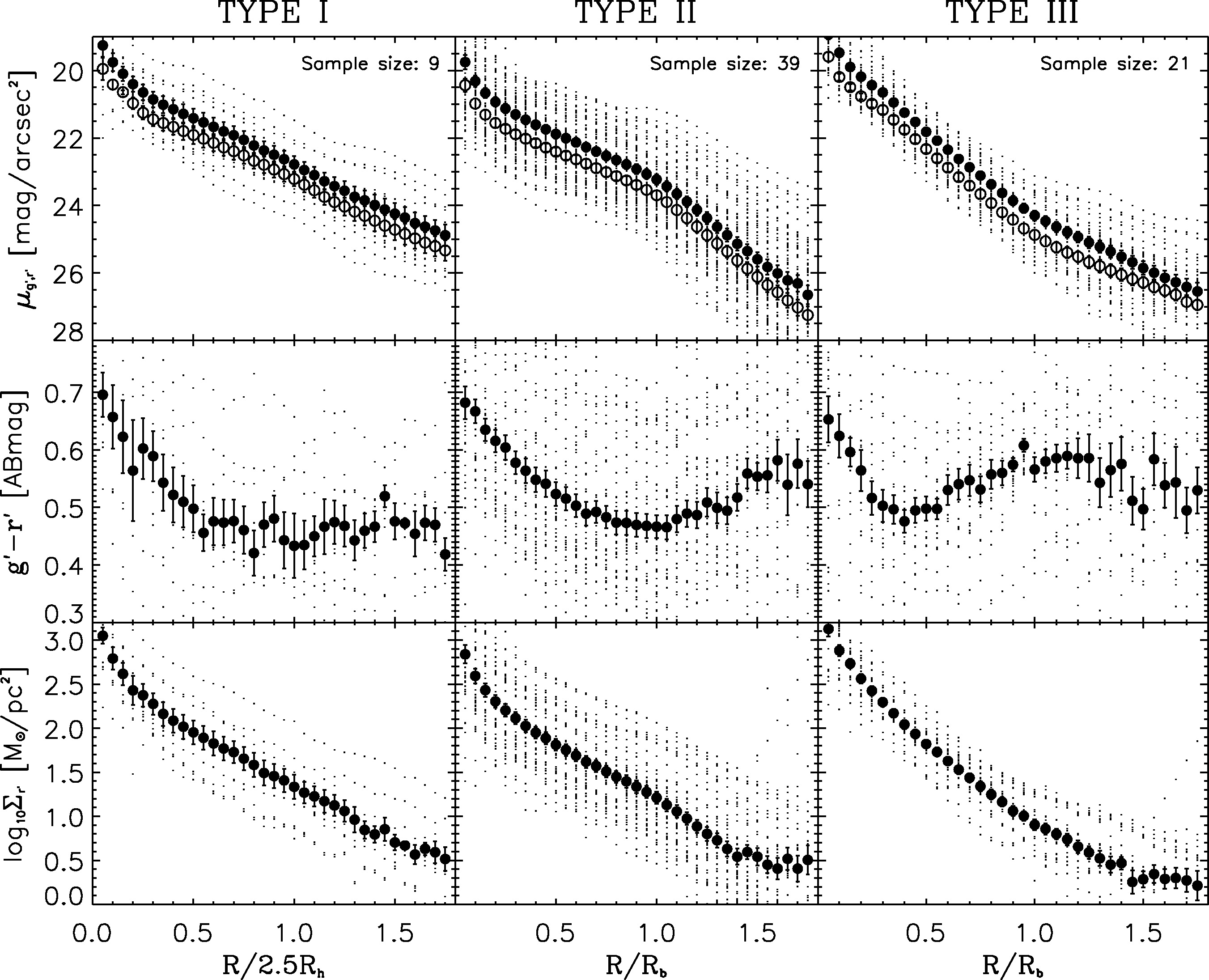}
\caption{Stacked profiles of 9 type~I ({\it left}), 39 type~II ({\it middle}) and
  21 type~III ({\it right}) galaxies.  The {\it top} row shows the $r'$ (filled
  circles) and $g'$ profiles (open circles).  The {\it middle} row shows the
  $g'-r'$ colour profiles.  The {\it bottom} row shows the computed mass
  density profiles.  Reproduced with permission from \citet{bakos+08}
}
\label{fig:bakos08} 
\end{figure}

Because of the age-metallicity degeneracy (e.g., \citealt{worthey94}),
colour profiles\index{colour profiles} are less constraining in
probing the stellar populations of outer disks.  \citet{azzollini+08}
stacked colour (approximating rest-frame $u-g$) profiles of different
profile types to $z \sim 1.1$.  They scaled the radius by the break
radius in type~II and III galaxies and to $2\,\Rd$ in type~I galaxies.
In the type~I galaxies, the resulting colour profiles are flat or
slightly rising, with no features.  Likewise the type~III galaxies
showed a range of gradients ranging from negative to positive with
increasing redshift.  Type~II galaxies instead become increasingly red
past the break at all redshifts.  For a sample of nearby late-type
galaxies, \citet{bakos+08} carried out a similar stacking analysis,
shown in Fig.~\ref{fig:bakos08}; they used the colour profiles to
obtain mass-to-light ratios, and thereby recovered the mass profiles.
They found that galaxies with type~II profiles have much weaker breaks
in the overall {\it mass} distribution than in the light.
\citet{zhengz+15} stacked a larger sample of 698 galaxies spanning a
wide range of mass and morphology.  In the bluer bands, in the
majority of galaxies they found type~II profiles, becoming type~I in
red bands, in contradiction to \citet{martin-navarro+12}, who found
numerous examples of type~II profiles in red bands, albeit weaker than
in the blue.  However, the latter authors stacked galaxies at fixed
$r_{90}$ rather than to the break radius, which probably smears any
weak breaks.

\subsection{Synthesis and Outlook}

Evidence that the outer disks of type~II profiles are due to migration
comes from the increasing age of stars past the break; in contrast an
outer disk dominated by accretion would have a constant age profile.
The coincidence of star formation breaks and stellar continuum breaks,
and the common break for stars of all ages, favours the churning
model.  However, the evidence is not yet conclusive.  Measurements of
kinematics in the outer disks would provide definitive proof of
whether migration is dominated by scattering or by churning.  Further
observational study and modelling of the role of environment is also
required.  The extent to which the mass profile deviates from pure
exponential needs further study and may help constrain the relative
importance of churning and scattering.


\section{Type~I Profiles}
\label{sec:typeI}

\begin{figure}[h!]
  \sidecaption
\includegraphics[angle=0.,width=0.6\hsize]{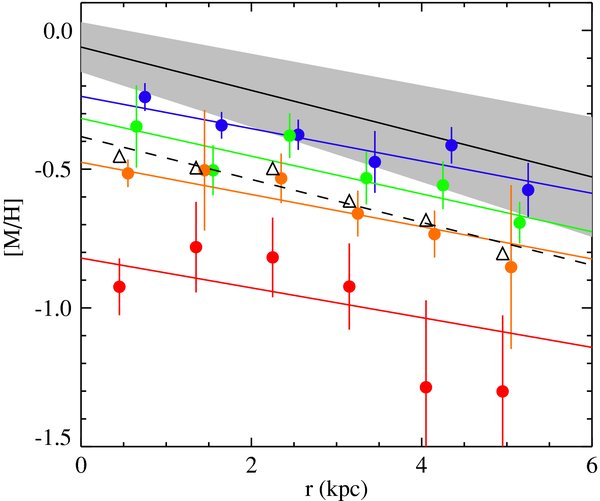}
\caption{The metallicity profiles of NGC~300.  Solid black line with
  grey shaded region: metallicity profile of 24 young A-type
  supergiants from \citet{kudritzki+08}.  Blue, green, orange and red
  circles are the metallicity profiles for stars with ages $<
  100\Myr$, $1-5\,\Gyr$, $5-10\,\Gyr$, and $10-14\,\Gyr$, respectively.  The
  black triangles and dashed line indicate the mean metallicity for
  the entire population.  Reproduced with permission from
  \citet{gogarten+10}  }
\label{fig:gogarten10} 
\end{figure}

Type~I profiles\index{type~I profile}, extending to $\sim 6-8\,\Rd$, are present in roughly
$10\%-15\%$ of field late-type spirals (\citealt{gutierrez+11}).  Examples
of galaxies with type~I profiles are sufficiently nearby that their
stellar populations can be resolved and studied directly.
NGC~300\index{NGC 300} is a relatively isolated, unbarred galaxy with
a type~I profile extending unbroken to 10 disk scale-lengths ($\sim
14\,\kpc$; \citealt{bland-hawthorn+05}).  Using {\it HST} to resolve the
stellar populations of the inner $\sim 5 \,\kpc$, \citet{gogarten+10}
showed that the majority of its stars are old (see
  also \citealt{vlajic+09}), with $80\%$ of them older than 6 Gyr.
\Rd\ increases from $1.1 \,\kpc$ for old stars to $1.3 \,\kpc$ for young
stars, indicating a modest inside-out formation, which is also evident
in the broad-band colours (\citealt{munoz-mateos+07}).
\citet{bland-hawthorn+05} estimated that everywhere beyond $\sim 6\,\kpc$ the
Toomre-$Q = 5 \pm 2$.  Thus spiral structure in NGC~300 is unlikely to
be strong, which, together with its low surface mass density, makes
churning inefficient, as confirmed by simulations (\citealt{gogarten+10}).
In agreement with this prediction, both young stars
(\citealt{kudritzki+08}) and old stars (\citealt{vlajic+09, gogarten+10})
exhibit the same negative metallicity gradient outside $2\,\kpc$, which
has remained quite constant over the past $\sim 10\,\Gyr$
(\citealt{gogarten+10}), as seen in Fig.~\ref{fig:gogarten10}.

NGC~300 is not unique: NGC~2403 is another nearby, isolated bulgeless
galaxy with a type~I profile.  Whereas NGC~300 has a strong \hi\ warp,
NGC~2403 has no warp (\citealt{bwilliams+13}).  All stellar populations,
including the young stars, follow the same density profile, with no
break.  The star formation histories are parallel at all radii to
$11\,\Rd$, with the surface density of star formation following the same
exponential profile as the overall density profile
(\citealt{bwilliams+13}).

\subsection{Origin of Type~I Profiles in Isolated Galaxies}

These observational results help constrain the origin of type~I
profiles.  \citet{minchev+11} proposed that the type~I profile of
NGC~300 was produced by very efficient migration driven by resonance
overlap scattering.  The presence of the radial metallicity gradients,
which have remained almost constant over $\sim 10\,\Gyr$, severely
limits the possibility that such migration has occurred in NGC~300.
The fact that its stars are predominantly old means that migration
would have had ample time to flatten any metallicity gradient if it
had been important.  The low mass and large Toomre-$Q$ of the disk
make it an unlikely candidate for strong spiral structure, making
extreme migration even less likely to have occurred.  The large
Toomre-$Q$ could have been produced by a bar heating the disk, but
NGC~300 is bulgeless, whereas angular momentum redistribution by a bar
would have produced a bulge-like central mass concentration
(\citealt{debattista+06}).  A similar case can be made for NGC~2403.

\citet{herpich+15} proposed that the shapes of density profiles depend
on the angular momentum of the gas corona\index{gas angular momentum}
out of which the disk forms.  At low spin parameter, $\lambda$, their
simulations formed type~III profiles (see below) while at high
$\lambda$ they formed galaxies with type~II profiles.  For galaxies
with an intermediate angular momentum ($0.035 \leq \lambda \leq 0.04$)
they formed type~I profiles.  This model very naturally accounts for
the relative rarity of the type~I profile.  Furthermore, because it
does not require stars to migrate, this model is consistent with the
metallicity gradients in NGC~300 and in NGC~2403.  The absence of a
break in the profiles of stars younger than $200 \Myr$ in NGC~300 also
supports this model.  Lastly, this scenario is also consistent with the
fact that the {\it inner} profiles of type~II and type~I galaxies are
consistent with each other (\citealt{gutierrez+11}).

\subsection{Type~I Profiles in Cluster Lenticulars\index{cluster lenticulars}}

\begin{figure}[h!]
\centering{
\includegraphics[angle=0.,width=0.7\textwidth]{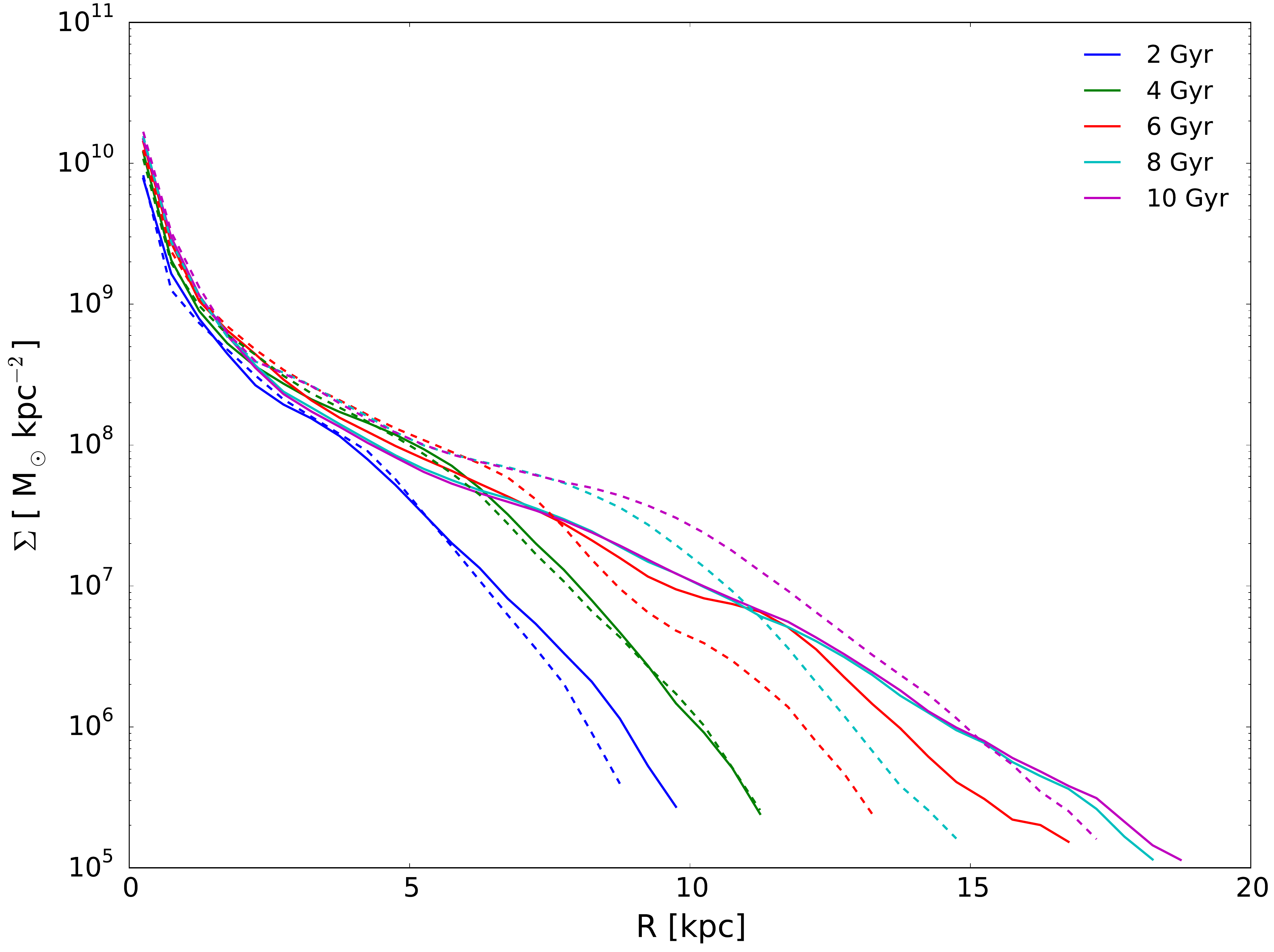}
\includegraphics[angle=0.,width=0.7\textwidth]{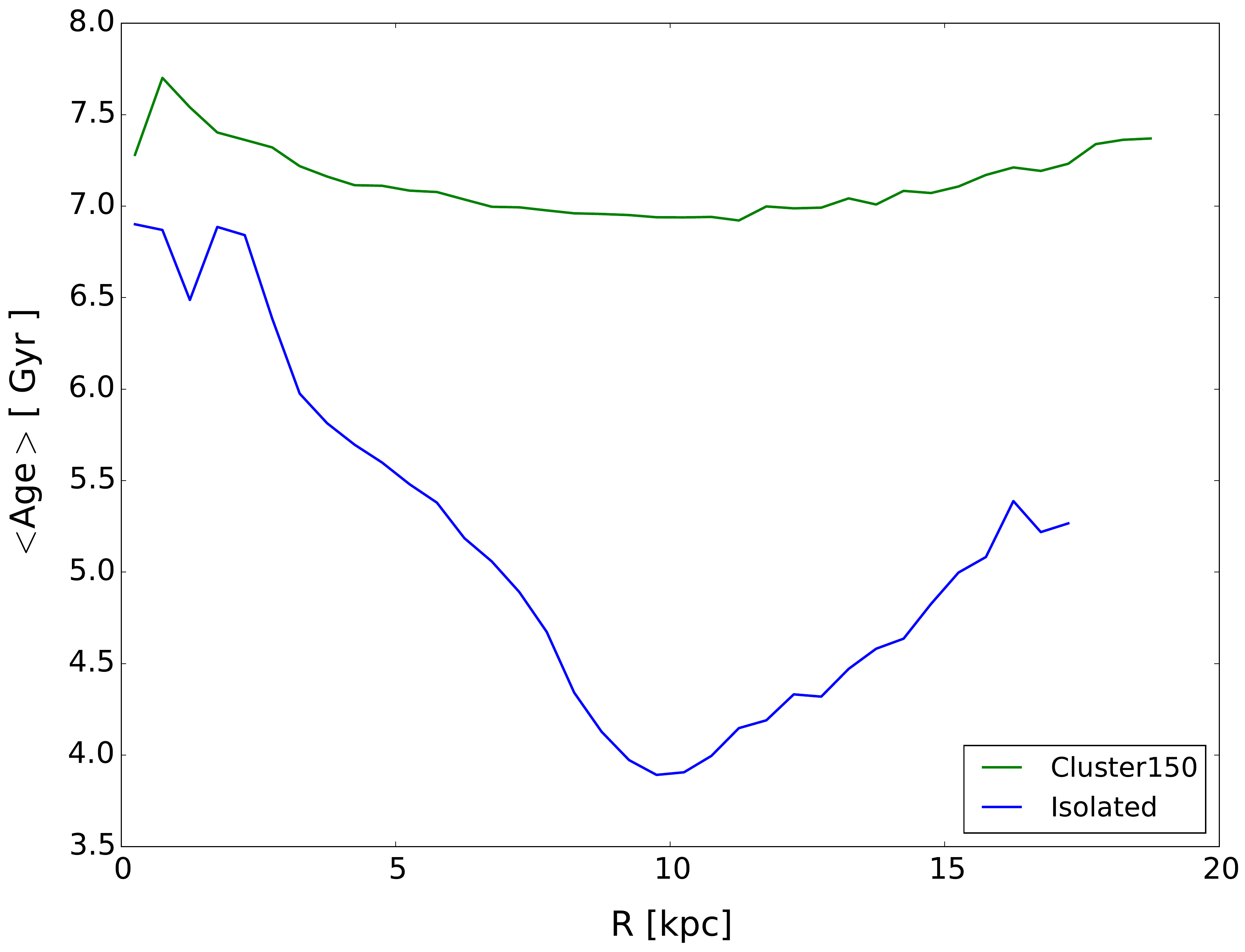}
}
\caption{{\it Top}: Evolution of surface density profile of the model
  of \citet{roskar+08a}, when evolved in isolation (dashed lines) and
  when it falls into a cluster, reaching pericentre at $5 \,\Gyr$ (solid
  lines).  Whereas the isolated simulation develops a type~II profile,
  the same system evolving in the cluster is transformed into a type~I
  profile via the joint action of ram pressure stripping and
  tidally-induced spirals.  {\it Bottom}: The final mean age profiles
  of the isolated and cluster models.  The isolated model has an
  increasing average age past the break, whereas the cluster model has
  a nearly flat age distribution.
  Reproduced with permission from \citet{clarke+16}  }
\label{fig:clarke16} 
\end{figure}

While type~I profiles are rare in the field, they are common amongst
cluster lenticulars (see Sect.~\ref{ssec:environment}). Since a
narrow range of $\lambda$s cannot account for a dominant population, a
different explanation for type~I profiles in cluster lenticulars is
required.
\citet{clarke+16} evolved the model of \citet{roskar+08a} (with $\lambda
= 0.065$) in a cluster environment.  Their model galaxy was quenched
by ram pressure stripping and transformed into a lenticular with very
little spiral structure at late times.  At the time of peak ram
pressure stripping, the disk had a break radius at $\sim 6\,\kpc$;
therefore the galaxy never gets a chance to form many stars beyond
this radius.  As shown in Fig.~\ref{fig:roskar08a}, the same model
evolved in isolation resulted in a type~II profile with a break at
$10\,\kpc$.  Nonetheless stars in the cluster galaxy moved to large
radii through migration via strong, tidally-induced,
spirals\index{tidally-induced spirals}, but not through heating of the
disk.  As a result, a type~I density profile developed, as shown by
the top panel of Fig.~\ref{fig:clarke16}.  This combination of
quenching by ram pressure stripping, and churning by tidally-induced
spirals accounts for the high incidence of type~I profiles amongst
cluster lenticulars.  Because of the strong churning into the outer
disk, \citet{clarke+16} predicted that the age profile of cluster
lenticulars with type~I profiles will be flat or slightly rising past
the point where the galaxy had a break radius before star formation is
quenched, as shown in the bottom panel of Fig.~\ref{fig:clarke16}.
Indeed for the Virgo Cluster, \citet{roediger+12} used colour profiles
to show that more than $70\%$ of galaxies with type~I profiles have
flat or slightly positive age gradients.

\section{Type~III Profiles}
\label{sec:typeIII}

\citet{erwin+05} identified two kinds of type~III profiles\index{type~III profile}.  The first
kind is produced by the excess light above an exponential from a
stellar halo\index{stellar halo}.  They argued that such galaxies can
be recognised by their outer isophotes becoming rounder.  M64 is an
example of such a galaxy (\citealt{gutierrez+11}); its outer profile is
well fit by a power law, typical of a halo.  The presence of
counter-rotating gas, with star formation concentrated within the
inner $4.5\,\kpc$, attests to M64 being a post-merger galaxy
(\citealt{watkins+16}), which probably produced the prominent halo.  In
such cases, \citet{martin-navarro+14} showed that the halo may mask the
presence of an underlying type~II profile in the disk, particularly at
low inclination.  \citet{erwin+05} estimated that haloes account for
$\sim 30\%$ of their sample of type~III profiles.

In the majority of their type~III galaxies, however, \citet{erwin+05}
found sharp transitions from one exponential to a shallower one, with
isophotes not becoming significantly rounder and, in some cases,
hosting spiral arms.  \citet{maltby+12b} showed that in $\sim 85\%$ of
cases the extrapolation of the bulge profile to large radii is
insufficient to account for the excess light at the break.  Thus,
except in cases where the light profile is flattened by the point
spread function (see the review by Knapen \& Trujillo, this volume),
the majority of type~III profiles are a disk phenomenon.

\subsection{Formation of Type~III Disk Profiles}

Breaks in type~III profile galaxies are not associated with a colour
change (\citealt{azzollini+08, bakos+08, zhengz+15}).  Changes in stellar
populations and star formation are therefore not the origin of these
profiles.

Anecdotally, \citet{erwin+05} noted that a number of galaxies with
type~III profiles exhibit asymmetries in the outer disks while
\citet{pohlen_trujillo06} found evidence of recent interactions amongst
type~III galaxies.  It is unsurprising therefore that many models of
type~III profile formation rely on mergers\index{mergers} and
interactions\index{interactions}.  \citet{laurikainen_salo01} used
simulations to show that interactions result in long-lasting shallow
outer profiles.  \citet{penarrubia+06} proposed that extended outer
disks are composed of tidally disrupted dwarf satellites accreted
directly onto the plane of the disk.  They showed that the degree of
rotational support of the accreted material depends on the satellite
orbit, with dispersion increasing with orbital eccentricity.  Instead
of the outer disk being accreted, \citet{younger+07} proposed that it
forms from the main disk itself after a minor merger forces gas to the
centre, causing the galaxy's inner region to contract and its outer
region to expand.  In the absence of gas, the pure $N$-body
simulations of \citet{kazantzidis+09} produced type~III profiles when
substructures\index{substructure} excited angular momentum
redistribution within the disk.  \citet{borlaff+14} demonstrated that,
in simulations, gas-rich major mergers, which destroy the main disk
then build a new one at large radius, develop into S0 galaxies with
type~III profiles.

Alternatively both \citet{minchev+12} and \citet{herpich+15b} proposed
that type~III profiles formed via the action of bars.
\citet{minchev+12} invoke resonance overlap in the presence of a bar
together with gas accretion to populate the outer disk.  In the model
of \citet{herpich+15b}, instead, stars in the inner disk gained energy
directly from the bar, getting boosted to increasingly eccentric
orbits, and reaching large radii.

Observations do not yet provide very strong constraints on how
type~III profiles form.  The lack of an observed difference between
type~III profiles in barred and unbarred galaxies (\citealt{borlaff+14,
  eliche-moral+15}), and the fact that type~III profiles are {\it less}
common amongst barred galaxies in the field (see Table~\ref{tab:demographics}) suggests that bars are unlikely to account for
the majority of type~III profiles.  In edge-on galaxies,
anti-truncations appear in thick disks\index{thick disks}, supporting
the view that a hot population is responsible for the outer disk
(\citealt{comeron+12}), contrary to the model of \citet{borlaff+14}.  The
most promising models therefore are the direct accretion model
(\citealt{penarrubia+06}) and the heating by substructure model
(\citealt{kazantzidis+09}), both of which produce hot outer disks.  The
model of \citet{younger+07} may also lead to hot outer disks because of
the rapid change of the inner potential.  Which of these models works
best can be decided by measuring the kinematics of both inner and
outer disks, as well as the chemistry of the outer disk.  Additional
constraints may be possible from comparing the properties of type~III
and type~I/II.  For instance, \citet{gutierrez+11} note that the inner
exponentials of type~III galaxies have shorter scale-lengths and higher
central surface brightness than galaxies with type~I profiles,
which argues against the direct accretion model but might favour the
model of \citet{younger+07}.


\section{Future Prospects}
\label{sec:future}

Much work remains to be done to better distinguish between competing
scenarios of outer disk formation, and to clarify what role, if any,
migration has played.  With the imminent launch of the {\it
  JWST}\index{JWST}, resolved stellar population studies will be
possible in a volume roughly three times larger than is accessible
with {\it HST}, improving the statistical robustness of age and
metallicity dissections of outer disks.  In addition, we identify
three areas that would greatly advance our understanding of disk
outskirts.
\begin{itemize}
\item
{\it Kinematics in the outer disks in type~II galaxies}
  The two classes of models for producing the outer disks of type~II
  profiles make very different kinematic predictions.  Models based on
  scattering of stars from the inner disk all produce radially hot
  outer disks.  In contrast, churning predicts a relatively cooler
  outer disk.  The observed line-of-sight velocity dispersion,
  \sig{\rm los}, is given by
  \begin{equation}
    \sig{\rm los}^2 = \frac{1}{2} \sin^2i\left[(\sig{\rm R}^2 + \sig{\phi}^2 + 2\sig{z}^2\cot^2i) - (\sig{\rm R}^2 - \sig{\phi}^2) \cos2\phi \right],
  \end{equation}
  where $\sig{\rm R}$, $\sig{\phi}$, and $\sig{z}$ are the velocity
  dispersions in the radial, tangential and vertical directions,
  respectively, $i$ is the galaxy's inclination and $\phi$ is the
  angle from the major axis of the disk in its intrinsic plane.  This
  can be written as
 $ \sig{\rm los}^2~=~a(R)~+~b(R) \cos2\phi.$
  The first term, $a(R)$, is positive definite and independent of
  $\phi$.  The second term, $b(R)$ is proportional to $-\beta_\phi =
  -(1-\sigma_\phi^2/\sigma_R^2)$; it is therefore negative in the
  inner disk where generally $\beta_\phi > 0$.  When the outer disk is
  formed of material that migrated via churning, then it is
  kinematically relatively cool and $b(R)$ remains negative.  If
  instead the outer disk is comprised of stars scattered outwards,
  then $\beta_\phi < 0$ and $b(R)$ is positive.  In the former case,
  $\sig{\rm los}$ peaks on the major axis, while it peaks on the minor
  axis for scattering.  This provides a very promising means for
  distinguishing which class of models is responsible for the outer
  disks in type~II galaxies.
  
  \item
  {\it Kinematics of type~III galaxies}
  Most models of type~III galaxies predict that outer disks, and, for
  some models, the inner disk, are hot, and that the outer disk is
  comprised of stars that have been moved out, or accreted directly.
  Confirming this prediction would give us an important insight into
  the properties of galaxies that have experienced these heating or
  accretion processes.
  
  \item
  {\it The outer disk of the Milky Way}
As with many other aspects of galaxy evolution, the Milky Way provides
an important laboratory for our understanding of disk outskirts.  In
the era of {\it Gaia}\index{Gaia} we should be able to study the ages,
metallicities, abundances and 3-D kinematics of stars past the break.
This will allow us to test the relations between the outer disk and
the inner disk, the warp, and the bar.  Accreted stars may also be
identified.  These data will result in a much more detailed picture of
the origin of the Milky Way's outer disk.
  
\end{itemize}
While the field of disk outskirts has made huge strides in the past
decade, the next decade promises to be even more exciting in terms of
understanding their assembly.  Confirmation of an important role for
radial migration would permit an important means for assessing its
{\it overall} action on disk galaxies.


\section{Acknowledgements}
VPD is supported by STFC Consolidated Grant \#ST/M000877/1.  We thank
Kathryn Daniel for providing us with the unpublished Figure
\ref{fig:daniel}.  Discussions with Peter Erwin have been especially
enlightening.  VPD thanks the Max-Planck-Institut f\"ur Astronomie for
hosting him during which time this review was started.

\bibliographystyle{spbasic}
\bibliography{allrefs}


\end{document}